\documentclass[aps,pra,showpacs,twoside,twocolumn,10pt]{revtex4-1}
\usepackage[colorlinks=true, citecolor=red, urlcolor=blue ]{hyperref}
\usepackage{times,epsfig,amssymb,amsfonts,amsmath,bm,subfigure,mathtools,amsthm,braket,soul,enumitem,color,physics,graphics,graphicx}
\usepackage[normalem]{ulem}
\usepackage{comment}
\usepackage{epstopdf}

\newcommand{\ph}[1]{{\color{black}#1}}

\begin{document}

\title{
Hybrid nonlocality via atom photon interactions with and without impurities
}

\author{Pritam Halder$^{1}$,  Ratul Banerjee$^{1}$, Saptarshi Roy$^{2}$, Aditi Sen(De)$^{1}$}

\affiliation{$^1$ Harish-Chandra Research Institute, A CI of Homi Bhabha National
Institute,  Chhatnag Road, Jhunsi, Allahabad - 211019, India\\
		$^2$ QICI Quantum Information and Computation Initiative, Department of Computer Science,
The University of Hong Kong, Pokfulam Road, Hong Kong	
	}

\begin{abstract}

To obtain Bell statistics from \textit{hybrid systems} composed of finite- and infinite-dimensional systems, we propose a \textit{hybrid measurement} scheme, in which the continuous mode is measured using the generalized pseudospin operators, while the finite (two)-dimensional system is measured in the usual Pauli basis.  
 Maximizing the Bell expression with these hybrid measurement settings leads to the  
 violations of local realism {in hybrid system} which is  referred to as \textit{hybrid nonlocality}. We demonstrate the utility of our strategy in a realistic setting of cavity quantum electrodynamics, where an atom interacts {resonantly} with a single mode of an electromagnetic field under the Jaynes-Cummings Hamiltonian. We dynamically compute the quenched averaged value of hybrid nonlocality in imperfect situations by incorporating disorder in the atom-cavity coupling strength. In the disordered case, we introduce two kinds of measurement scenarios to determine the Bell statistics -- in one situation, experimentalists can tune the optimal settings according to the interaction strength while such controlled power is absent in the other case.  In contrast to the oscillatory behavior observed in the ordered case, the quenched averaged violation saturates to a finite value in some parameter regimes in the former case.
 We also examine the connection between Wigner negativity and  hybrid nonlocality. 
\end{abstract}	

\maketitle

\section{Introduction}
\label{sec:intro}
John S. Bell discovered a set of  mathematical inequalities \cite{Bell1964}  to determine whether a theory can be compatible with the assumptions of locality and reality \cite{epr1935}. It was shown that there exist quantum states that violate Bell's inequalities, thereby revealing the incompatibility  with Einstein, Podolsky and Rosen's argument of local realism \cite{epr1935}.
 This non-classical feature of quantum mechanics is, in fact, one of the foundational pillars to \textit{understand} the quantum world. It has also been experimentally demonstrated multiple times over the past fifty years that states having quantum correlations {\it aka} entanglement, violate Bell inequalities \cite{bellexp1,bellexp2, bellexp3,bellexploophole}. Apart from fundamental implications on the structure of the theory, violation of Bell inequalities also certifies supremacy in a number of tasks like device-independent quantum key distribution \cite{diqkd1,diqkd2}, random number generation \cite{rng}, self-testing \cite{st} etc., as  a resource quantifier \cite{rt1,rt2} for developing quantum technologies.


It has recently been realized that the development of modern quantum architectures requires cross-platform information processing where different nodes of the network comprise different types of systems, commonly referred to as {\textit{hybrid systems}} \cite{hybrid1}
consisting of discrete and continuous variable (CV) counterparts \cite{hybrid1,hybrid2}. 
The majority of theoretical and experimental research of hybrid systems to date have centered on characteristics like nonclassicality \cite{hybridnc}, and entanglement \cite{hybrident,hybrident2,Arkhipov2018}.  It is important here to emphasize that investigating  violations of Bell inequality differs qualitatively  from studying correlations like entanglement since the statistics required to construct the Bell expression depend on the choice of measurements. For example, even the singlet state may  fail to show any violation if the appropriate measurements are not chosen to extract the Bell statistics. Our aim in this work is to construct a suitable framework for testing the violation of local realism in hybrid systems, thereby, illustrating the interplay between hybrid states and measurements.
\begin{figure}
\includegraphics[width=1.0\linewidth]{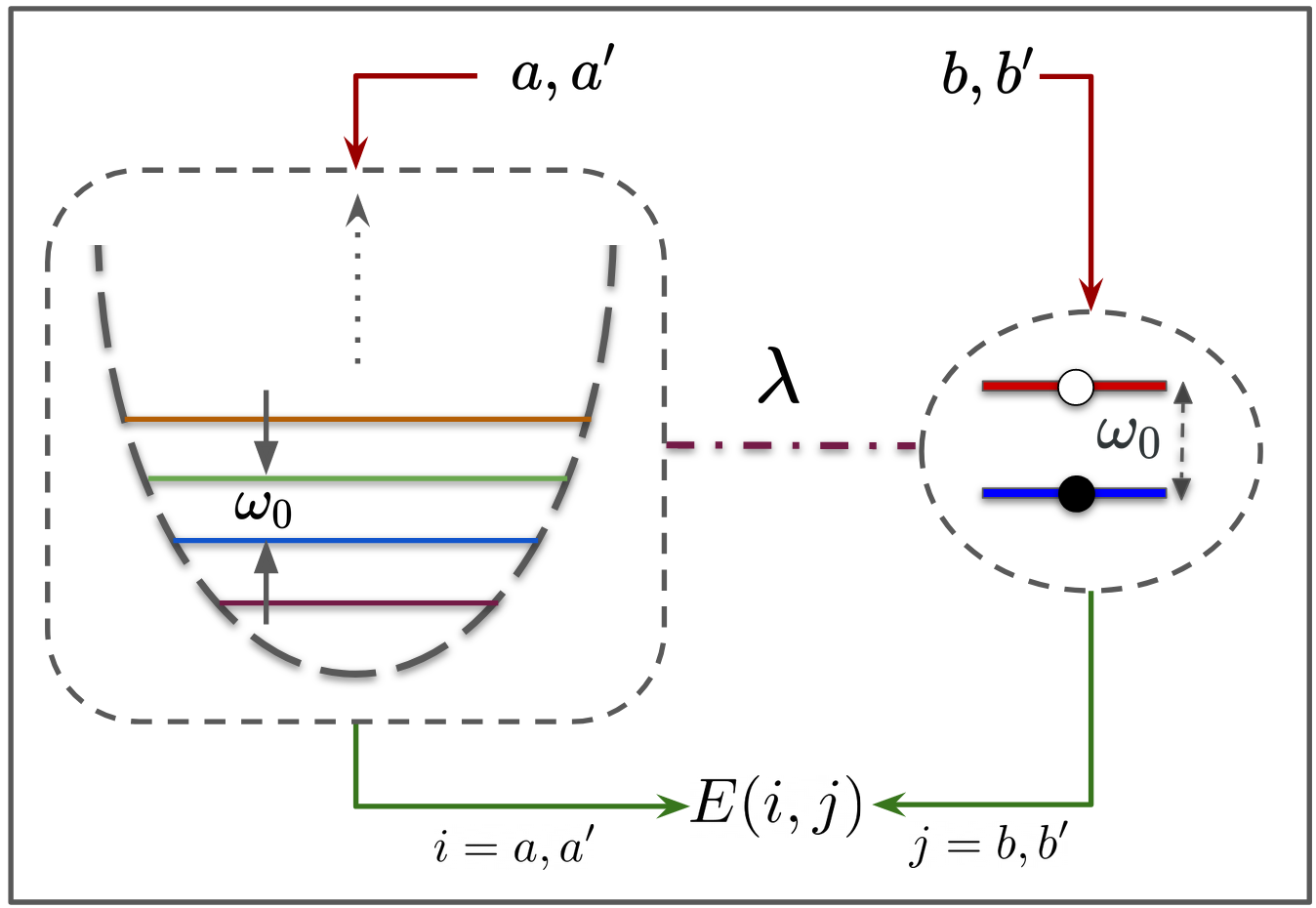}
\caption{(Color online.) Schematic diagram of the extraction of Bell statistics via correlation functions,  $E(i,j)$, from a hybrid system with $i = a,a^\prime$ and $j =b,b^\prime$ being the measurement settings at sites $A$ and $B$ of the entire system.
The hybrid system consists of a single mode of a quantized electromagnetic field (modelled as a cavity), referred to as subsystem $A$, and an atom (subsystem $B$) which interacts via Jaynes-Cummings Hamiltonian. 
$\omega_0$ is the resonant frequency which corresponds to both the frequency of the cavity and the energy gap of the atom. The atom-photon interaction follows a coupling strength $\lambda$.}
\label{fig:schematic} 
\end{figure}

In discrete systems, violations of Bell inequalities are routinely investigated both theoretically and experimentally, although, for CV systems, the Bell expression to examine the violation of local realism is not straightforward. Specifically, J.S. Bell \cite{Bell1964} himself argued that Gaussian states which possess positive (Gaussian) Wigner functions, would not violate these inequalities. This was believed to be true for a long time until it was disproved by employing  \cite{Banaszek1998, Chen2002} non-Gaussian measurements to extract the violation out of Gaussian states. In particular,  the maximal violation for the Einstein-Podolsky-Rosen (EPR) state \cite{epr1935} is obtained using pseudospin operators \cite{Chen2002}  (later generalized   pseudospin operators \cite{Zhang2011}). Nevertheless, the connection of negativity in Wigner function with its maximal  violation of Bell inequality by a quantum state is an interesting question and has been pursued in several works for CV systems \cite{Banaszek1998,Chen2002,sap2018}. We intend to investigate the link, if any, between these two quantities for hybrid quantum systems. 

To investigate the violation of Bell inequality for a prototypical hybrid system consisting of an atom (a qubit) and a single mode of light (CV system)  (see Fig. \ref{fig:schematic}), we integrate the concepts both from qubit and CV systems. Specifically,  to acquire the Bell statistics, we introduce a \ph{\textit{hybrid measurement}} strategy that uses Pauli measurement on an atom and generalized pseudospin operators in the CV component of the system \cite{Chen2002,Zhang2011}. \ph{We define \textit{hybrid nonlocality} to be the extraction of nonlocal correlation from the hybrid system with the aid of a hybrid measurement scheme.}

In this work, initial states are taken to be either a product or classically correlated or entangled in the atom-light bipartition which is allowed to interact via the Jaynes-Cummings Hamiltonian \cite{JC,Gerry}. In the case of separable states,  the CV system is either in the Fock or squeezed or coherent state. On the other hand, the hybrid cat state \cite{Arkhipov2018} is chosen to be the initial state which is, in general, entangled.
We find that the maximal violation of Bell inequality oscillates with time and is highly dependent on the parameters involved in CV systems. We establish a connection between the violation of local realism, initial entanglement, and hybrid Wigner negativity volume \cite{Karol2004,Arkhipov2018}. \ph{Note that, throughout the work, we assume that both the frequency of oscillation of the atom and the field are the same, i.e., the resonance condition, to simplify the complexity in the calculation.  However, the formalism of our work also applies to the detuned scenario  where the value  and time of oscillation of Bell inequality violation can change compared to the resonance condition depending on the detuning parameter.}

Departing from the ideal situation, we introduce impurities in the interaction strength.
 We demonstrate that the quenched averaged violation persists even in the presence of disorder,  the oscillations observed in the ordered system vanish and saturate to some nonvanishing value when it is assumed that experimentalists are permitted to adjust the measurement settings in each realization of disorder. 
 On the other hand, if the measurements are fixed to be optimal for the mean value of the random Gaussian disorder, we report that a violation of local realism is obtained only for a certain period of time, after which no violation is detected.

The paper is organized as follows. We introduce the formalism to evaluate hybrid nonlocality using hybrid measurements in Sec. \ref{sec:formalism}. The details of setting up the optimization problem are discussed in Sec. \ref{sec:maximization_bell}, while the Wigner negativity volume is defined in Sec. \ref{sec:formalismwigner}. The generation and dynamics of hybrid nonlocality via atom-photon interactions are illustrated in Sec. \ref{sec:3}. We provide a brief primer on the atom-photon interactions via the Jaynes-Cummings Hamiltonian in Sec. \ref{sec:JCham}, and examples of different initial states, separable in Sec. \ref{sec:initialstates} and entangled in Sec. \ref{sec:initialent}.
The violation obtained  in the dynamical states is compared with their Wigner negativity volume in Sec. \ref{sec:Wigner}. Finally, we study the effects of imperfections on the Bell-violation in Sec. \ref{sec:imperfection}, where we investigate the role of disorder in the coupling strength of the atom-photon interactions. A conclusion is provided in Sec. \ref{sec:conclusion}.

\section{Hybrid nonlocality using Hybrid measurements}
\label{sec:formalism}

Let us start our discussion with a particular variant of Bell  inequalities, namely the Clauser-Horn-Shimony-Holt (CHSH) inequalities \cite{chsh1,chsh2}. It turns out to be the facet (tight)  ones in the $2-2-2$-configuration ($2$ parties, $2$ measurement settings, $2$ outcomes per measurement) \cite{Fine1982}.

Let us explain the CHSH inequality briefly which gives us the premise to detect quantum correlations in hybrid systems.  Consider an arbitrary two-party state, $\rho$, shared between  $A$ and $B$ in spatially separated locations. For each party, two different dichotomic measurements, $a, a^\prime$ and $b, b^\prime$ are employed. Depending on the settings, the measurement outcomes are denoted by $A_{i}$ and $B_{j}$  with $i = a, a^\prime$ and $j = b, b^\prime$ for $A$ and $B$ respectively.
The assumptions of locality and realism lead to the upper bound $2$ of the Bell expression, given by
\begin{eqnarray}
 |\mathcal{B}| =  |E(a,b) + E(a^\prime, b) + E(a, b^\prime) - E(a^\prime, b^{\prime})|,
\label{eq:bellexp}
\end{eqnarray}
where the correlation function $E(i,j) = \langle A_i B_j \rangle_{\rho} $.
 The maximal violation that can be achieved via a quantum state is given by the Tsirelson's bound \cite{Cirelson1980} for which $|\mathcal{B}^Q|_{\max} = 2\sqrt{2}.$

The statistics of the Bell CHSH expression are generated by the outcomes of dichotomic measurements on quantum states $\rho$. For continuous variable systems, the choice of these two outcome measurements that results in maximal violation,  in principle, requires a search over an infinite number of parameters. This, in general, makes the problem of choosing optimal  measurements very hard  for CV systems. \textcolor{black}{For a long time, following Bell's stipulation \cite{Bell2004}, it was believed that Gaussian states could not demonstrate a violation of Bell inequalities. Contrasting, it was pointed out  \cite{Banaszek1998} that using non-Gaussian measurements, one can extract Bell-violating statistics from a Gaussian state. However, such measurements could not saturate the Tsirelson's bound even for an EPR state. A few years later, a type of dichotomic continuous variable pseudospin measurements was introduced \cite{Chen2002} that was shown to achieve the Tsirelson's bound for the EPR state. From then on, the pseudospin measurements and their generalized versions were routinely used in the literature \cite{Zhang2011,sap2018} and we also use them to construct Bell inequalities for the hybrid systems. Variants of the pseudospin measurements have also been implemented experimentally \cite{Parity-exp1}.}
The generalized pseudospin operators \cite{Chen2002,Zhang2011,sap2018} are defined by,
\begin{eqnarray}
S_z^q &=& \sum_{\substack{n=0 \\ 2n +q \geq 0}}^\infty |2n +q+1\rangle\langle 2n + q +1| - |2n +q\rangle\langle 2n + q |, \nonumber \\
S_-^q &=& \sum_{\substack{n=0 \\ 2n +q \geq 0}}^\infty |2n +q\rangle\langle 2n + q +1| = (S_+^q)^\dagger,
\label{eq:gen_pseudo_spin}
\end{eqnarray}
where $q$ is an integer. \textcolor{black}{ For a brief survey of the potential experimental implementability of these measurements, see Appendix. \ref{app:proposal}.
}

\subsection{Constructing Bell inequalities using hybrid measurement strategies} 
\label{sec:maximization_bell}
We are now positioned to formulate the Bell CHSH expression of a hybrid quantum system. While the party with the qubit system performs the usual Pauli measurements, the party which possesses the infinite dimensional system employs the pseudospin measurements to extract the required statistics for constructing the Bell expression.
The correlation functions of a given state, $\rho$, that constitute the Bell expression can be written as
\begin{eqnarray}
    E(i,j) = \text{Tr} [S_i^q \otimes  \sigma_j \rho ],
\end{eqnarray}
where \ph{$\sigma_j = \cos \theta_j \sigma_z + \sin \theta_j (e^{-i\phi_j}\sigma_+ + e^{i\phi_j}\sigma_-)$}, and $S^q_i = \cos \theta_i S_z^q + \sin \theta_i (e^{i\phi_i}S_+^q + e^{-i\phi_i}S_-^q)$. The Pauli ladder operators can be written in terms of Pauli spin operators as $\sigma_{\pm} = (\sigma_x \pm i \sigma_y)/2$. Similarly, like the usual Pauli operators, $S_x^q$ and $S_y^q$ are given by $S_+^q + S_-^q$ and $-i(S_+^q - S_-^q)$ respectively, where $S_+^q$ and $S_-^q$ are given in Eq. \eqref{eq:gen_pseudo_spin}. The measurements are parameterized by their respective polar $(0 \leq \theta_k \leq \pi)$ and azimuthal $(0 \leq \phi_l \leq 2\pi)$ angles. We are interested in the maximal violation of Bell inequality by a given quantum state which we obtain by optimizing over the measurement parameters as
\begin{eqnarray}
    |\mathcal{B}|_{\max} = \max_{\bm{\theta},\bm{\Phi},q} |\mathcal{B}|, 
\end{eqnarray}
where $\bm{\theta} = (\theta_a,\theta_{a^\prime},\theta_b,\theta_{b^\prime})$ and $\bm{\Phi} = (\phi_a,\phi_{a^\prime},\phi_b,\phi_{b^\prime})$. \\
\ph{
It is important to notice that the above formalism can be generalized to a multipartite setting also. In particular, instead of the CHSH inequality, one can also apply hybrid measurement strategies which consists of pseudospin measurements in local CV systems and Pauli basis measurements in qubit subsystems to construct Mermin's inequality \cite{Mermin1990,DEFABRITIIS2023100177}.
An effectiveness of these inequalities can be tested by considering systems consisting of multiple atoms and single-mode cavities described by the Dicke model \cite{Dicke1954} or the Jaynes-Cummings-Hubbard model \cite{Hartmann2006,Greentree2006,Angelakis2007} which describes the dynamics of multiple coupled single-mode cavities, each with a two-level atom. }

\subsubsection{Maximization of Bell expression}
\label{sec:max}
We now present a brief primer on how to obtain  $|\mathcal{B}|_{\max}$ by performing maximization over the settings, i.e., $\bm{\theta},\bm{\Phi}$ and $q$.
The elements of the correlation matrix, $T$, in the hybrid systems is defined by $T_{kl} = \langle S_k^{q}  \otimes \sigma_l  \rangle$, where $k,l = x,y,z$.
Following the prescribed strategy \cite{horodecki1995}, the maximal violation of Bell inequality can again be expressed as $\ph{|\mathcal{B}|_{\max}=}2 \sqrt{\Lambda_1 + \Lambda_2}$, where $\Lambda_1$ and $\Lambda_2$ denote the two largest eigenvalues of $T^\dagger T$. 
Although this nine parameter problem is hard to track  analytically, we can reduce the number of  parameters to be optimized for some cases and can obtain an analytical closed form expression of the optimal Bell-CHSH violation. \ph{Moreover, note that, for states of the form $\ket\psi = \sum_{n=0}^\infty \ket{n+k}\ket{n}$ where the difference in photon number between two modes is $k$,  no violation is observed when $k$ is odd and the maximal violation is observed when $q=k$ \cite{Zhang2011}. However, in this work, we have hybrid states (Eq. \ref{eq:psi_t}), where the difference in photon number between two modes increases with the increase of photon number in the CV mode. Therefore, it is very difficult to track down the optimal value of $q$ in a systematic way. Throughout the work, we choose the optimal value of $q$ either analytically or  numerically within a fixed range of $q$ values.}

\subsection{Wigner negativity volume}
   \label{sec:formalismwigner}
Introduced by Wigner in 1932 \cite{Wigner1932}, Wigner functions have been rigorously used in quantum optics as a detector of non-classicality \cite{Karol2004}. The Wigner function is defined in the phase space as the Fourier transform of the symmetrized characteristic function \cite{adesso2014}. 
For a single mode CV system, $\rho$, the Wigner function can be expressed as \cite{Gerry}
\begin{eqnarray}
    W(x,p) = \frac{1}{2\pi}\int_{-\infty}^{+\infty} \bra{x+\frac{1}{2}q}\rho \ket{x-\frac{1}{2}q} e^{ipq} dq,
\end{eqnarray}
where $\ket{x\pm\frac{1}{2}q}$ denote the eigenstates of the $X$-quadrature. Although it satisfies the normalization condition, i.e., $\int W(x,p) dx dp = 1$, it is interpreted as a quasi-probability distribution since the Wigner function may not always be positive throughout the phase space. The negativity of the Wigner function $(W(x,p) < 0)$ for some states, in turn, aids to gauge its nonclassicality. The nonclassicality detection capabilities of the Wigner function can be turned into a measure when one considers the volume of the Wigner function in the phase space where it is negative. Technically, it is called the Wigner negativity volume \cite{Karol2004}, and is defined as
\begin{eqnarray}
  \mathcal{V}_n = \int_{W(x,p)<0} W(x,p) dxdp. 
\end{eqnarray}
We now briefly describe the prescription for the computation of Wigner negativity volume for hybrid quantum systems  \cite{Arkhipov2018}. The Wigner function of a hybrid quantum state $\rho$ consisting of both discrete and continuous variables is given by \cite{Nemoto2016}
\begin{eqnarray}
    W(\theta, \phi, \beta) = \text{Tr}\big[\rho  ~\hat{\Delta}_c(\beta)\otimes \hat{\Delta}_d(\theta, \phi)\big],  
    \label{eq:hybridwigner}
\end{eqnarray}
where 
\begin{eqnarray}
    \hat{\Delta}_c(\beta)=\frac{2}{\pi}\hat{D}\hat{\Pi}_c\hat{D}^\dagger, \hspace{0.2cm} \text{and} \hspace{0.2cm} \hat{\Delta}_d(\theta, \phi)=\frac{1}{2}\hat{U}\hat{\Pi}_q \hat{U}^\dagger,
    \label{eq:kernel}
\end{eqnarray}
stand for kernel operators corresponding to the CV and qubit systems, respectively. Here, $\hat{D}(\beta) = e^{\hat{a}^\dagger \beta - \hat{a}{\beta^*}}$ is the displacement operator and $\hat{\Pi}_c=e^{i \pi \hat{a}^\dagger \hat{a}}$ is the parity operator for the CV system. In the qubit scenario, $\hat{U}\in SU(2) $, which can be written in terms of Pauli operators $\sigma_i, i=1,2,3$ as $\hat{U}=e^{i\sigma_3\phi} e^{i\sigma_2\theta} e^{i\sigma_3\Phi}$, where $\theta\in [0,\pi]$ and  $\phi,\Phi\in[0,2\pi]$ and the parity operator for qubit is given by $\hat{\Pi}_d = \mathbb{I}-\sqrt{3}\sigma_3$. 

The normalization condition for the hybrid Wigner function is obtained from the Haar measure \cite{bengtsson_zyczkowski_2006}, $d \Omega$, such that $\int W d\Omega  = 1$. The integral measure $d\Omega$ is a product of the Haar measure for qubits $(d\nu)$ which turns out to be the differential volume of the SU$(2)$ space corresponding to a qubit and the differential volume of the field of coherent states $d^2 \beta$. The integral measure, therefore, becomes
\begin{eqnarray}
    d\Omega = d\nu d^2\beta = \frac{1}{\pi}\sin{2\theta}d\theta d\phi d^2\beta,
    \label{eq:measure}
\end{eqnarray}
with $\theta \in [0,\pi/2]$ and $\phi \in [0,2\pi]$. We then compute the hybrid Wigner negative volume ($\mathcal{V}_n$) by 
\begin{eqnarray}
    \mathcal{V}_n = \frac{1}{2}\int(|W|-W) d\Omega = \frac{1}{2}\bigg(\int|W|d\Omega-1\bigg).
    \label{eq:negvol}
\end{eqnarray}
In this work, we intend to investigate whether there exists some connection between the maximal violation of hybrid Bell inequality, $|\mathcal{B}|_{\max}$ and Wigner negativity volume, $\mathcal{V}_n$.

\section{Generation and dynamics of hybrid nonlocality}
\label{sec:3}
We explore how atom-photon interactions can generate hybrid  correlations from an initially separable state, which can lead to the violation of the Bell-CHSH inequality based on pseudospin operator formalism  as laid out in Sec. \ref{sec:maximization_bell}. Further, starting from a hybrid entangled initial state that violates the Bell inequality, we also track how the dynamics generated by the atom-photon interactions can sustain, enhance or deteriorate the initial violation. Before moving on to explicit examples, we first present a brief description of the atom-photon interaction that is considered in this manuscript.

\subsection{Modelling atom-photon interactions: Jaynes-Cummings Hamiltonian}
\label{sec:JCham}
Interaction between quantized electromagnetic field and atomic system that can be modelled via the prototypical Jaynes-Cummings (JC) Hamiltonian. Here the light and  matter are both taken as quantum,i.e., a quantized electromagnetic field and a two level system respectively. Recently, it has been possible to create environments capable of supporting a single mode field, such as small microwave cavities, optical cavities etc \cite{cavity_qi1,cavity_qi2,cavity_qi3,MEYSTRE1992243}. In our case, we will concentrate on a setup where a single mode light field interacts with a two level system.
The atomic transition and inversion operators obeying Pauli spin algebra  are respectively given by
\begin{eqnarray}
    {\sigma}_{+} = \ket{e}\bra{g}, \hspace{0.1cm} {\sigma}_{-} = \ket{g}\bra{e},\hspace{0.1cm}\text{and}\hspace{0.1cm} {\sigma}_{z} = \ket{e}\bra{e} - \ket{g}\bra{g}.
\end{eqnarray} 
For all these operators, we have $[{\sigma}_{+},{\sigma}_{-}] = {\sigma}_{z}$, $[{\sigma}_{z},{\sigma}_{\pm}] = \pm 2 {\sigma}_{\pm}$.
The total atom-field Hamiltonian can be written as
\begin{eqnarray}
\nonumber
    {H} &=& {H}_{A} +{H}_{F}+{H}_{I} \\
            &=& \frac{1}{2}\hbar\omega_{0}{\sigma_{z}} + \hbar\omega{a}^{\dagger}{a} +\hbar \lambda
    ({a} + {a}^{\dagger})( \hat{\sigma}_{+} +  {\sigma}_{-} ).\\\nonumber
\end{eqnarray}
Here $H_A$ is the atomic Hamiltonian with energy gap $\hbar \omega_0$ between the  energy levels ($\hbar$ is Planck's constant), $H_F$ describes the free electromagnetic field with frequency $\omega$ and $H_I$ characterizes the interaction between atom and field with interaction strength $\lambda$. Note that the terms ${a}^{\dagger}{\sigma}_{+}$ and ${a}{\sigma}_{-}$ correspond to the energy non-conserving terms which vary rapidly in time, are dropped using rotating wave approximation (RWA). Finally, we can write the JC Hamiltonian as
\begin{eqnarray}
   {H} = \frac{1}{2}\hbar\omega_{0}{\sigma_{z}} + \hbar\omega\hat{a}^{\dagger}{a} +\hbar \lambda ( {a}{\sigma}_{+} + {a}^{\dagger}{\sigma}_{-} ).
\end{eqnarray}
In our work, we take the detuning parameter $\delta = \omega - \omega_0=0$, i.e., the resonance condition, $\omega = \omega_0$ \ph{for simplicity}. Therefore, our JC \ph{ Hamiltonian} becomes
\begin{eqnarray}
    H = H_0 + H_I,
        \label{eq:jc_hamiltonian}
\end{eqnarray}
where 
\begin{eqnarray}
\nonumber
    H_0 &=& \frac{1}{2}\hbar\omega_{0}{\sigma_{z}} + \hbar\omega_0{a}^{\dagger}{a}, \\ 
\text{and   }    H_I &=& \hbar \lambda ( {a}{\sigma}_{+} + {a}^{\dagger}{\sigma}_{-}  ).
\end{eqnarray}
In the interaction picture, the time evolution unitary of Eq. (\ref{eq:jc_hamiltonian}) can be written as 
\begin{eqnarray}
\nonumber
    U_{\lambda}(t) &=&e^{-i{H}_{I} t/\hbar} = e^{-i\lambda t
    ({a}{\sigma}_{+} + {a}^{\dagger} {\sigma}_{-} )} \\\nonumber &=&\cos{(\sqrt{a^\dagger a} \lambda t)}\ket{e}\bra{e} - i a \frac{\sin{(\sqrt{a^\dagger a}\lambda t)}}{\sqrt{a^\dagger a}}\ket{e}\bra{g} \\ \nonumber
    &-& i a^\dagger \frac{\sin{(\sqrt{a a^\dagger}\lambda t)}}{\sqrt{a a^\dagger}}\ket{g}\bra{e} + \cos{(\sqrt{a^\dagger a}\lambda t)}\ket{g}\bra{g}. \\ 
\end{eqnarray}
The above unitary operator will be used to obtain dynamical state whose nonlocal features will be studied. \ph{In the subsequent sections, we analyze how hybrid nonlocality emerges in various initial states, commonly appearing in the domain of quantum optics.}

\subsection{Generation of hybrid nonlocality  for different initial separable states}
\label{sec:initialstates}

Let us analyze how nonlocality can be generated with time via atom-photon interactions from  initial separable hybrid states. Let us take the initial state as any general single qubit and a single mode continuous variable state in the eigenbasis of their respective local Hamiltonians. It can be expressed as
\begin{eqnarray}
\nonumber
 \ket{\psi(0)}_{\text{atom}} = C_{g}\ket{g} + C_{e}\ket{e}, \text{ and }\ket{\psi(0)}_{\text{field}} = \sum_{n=0}^{\infty}C_{n}\ket{n}.\\
   \label{eq:singlestates}
\end{eqnarray}
Clearly, $|g(e)\rangle$ denotes the ground (excited) state of $H_A$ wheras $\{\ket{n}\}$ constitute the Fock basis. Therefore, any separable hybrid state can be written as a convex combination of states written as a tensor product of qubit and CV states of the form given in Eq. \eqref{eq:singlestates}.

 \subsubsection{Initial product hybrid states}
 The initial hybrid state in product form can, in general, be written  as $\ket{\psi(0)} = \ket{\psi(0)}_{\text{field}} \otimes \ket{\psi(0)}_{\text{atom}}$. As the initial state is product in field-atom bipartition, following unitary with the JC interaction, the time evolved hybrid state can be written as
 \begin{eqnarray}
 \nonumber
    &&\ket{\psi(t)}= U_{I}(t)\ket{\psi(0)}\\ \nonumber &&= \sum_{n=0}^{\infty}\big\{[C_{e}C_{n}\cos{(\lambda t \sqrt{n+1})} - i C_{g}C_{n+1}\sin{(\lambda t \sqrt{n+1})}]\ket{e} \\ && +  [-iC_{e}C_{n-1}\sin{(\lambda t \sqrt{n})}  + C_{g}C_{n}\cos{(\lambda t \sqrt{n})}]\ket{g}\big\}\ket{n}. 
    \label{eq:u_psi}
\end{eqnarray}
Let us assume that the atom is initially in the excited state, i.e., $C_e=1$, $C_g=0$ and we write Eq. (\ref{eq:u_psi}) as
\begin{eqnarray}
    \ket{\psi(t)}= \ket{\psi_{g}(t)}\ket{g} + \ket{\psi_{e}(t)}\ket{e},
    \label{eq:psi_t}
\end{eqnarray}
where
\begin{eqnarray}
\nonumber
     \ket{\psi_{g}(t)} &=& -i\sum_{n=0}^{\infty}C_{n}\sin(\lambda t \sqrt{n+1})\ket{n+1}, \\ 
     \text{and}
    \ket{\psi_{e}(t)} &=& \sum_{n=0}^{\infty}C_{n}\cos(\lambda t \sqrt{n+1})\ket{n}.
\end{eqnarray}
The correlation matrix $T_{ij}$ as described in Sec. \ref{sec:max} 
for the states of the form in Eq. \eqref{eq:psi_t} can be computed as
\begin{eqnarray}
\nonumber
     &&T_{ij} = \bra{\psi(t)}S_i^q \otimes \sigma_{j}   \ket{\psi(t)}\\\nonumber&&=\bra{\psi_g(t)}S_i^q\ket{\psi_g(t)}\bra{g}\sigma_j\ket{g} + \bra{\psi_e(t)}S_i^q\ket{\psi_e(t)}\bra{e}\sigma_j\ket{e}\\\nonumber &&+\bra{\psi_g(t)}S_i^q\ket{\psi_e(t)}\bra{g}\sigma_j\ket{e} + \bra{\psi_e(t)}S_i^q\ket{\psi_g(t)}\bra{e}\sigma_j\ket{g}.\\
\end{eqnarray}
We now compute the maximal violation of the Bell-CHSH inequality, $|\mathcal{B}|_{\max}$ by taking different initial states, i.e., different $C_n$'s at any instant of time.

\paragraph{$k{\text{th}}$ Fock basis as the initial field state.}

\begin{figure}
\includegraphics[width=1.0\linewidth]{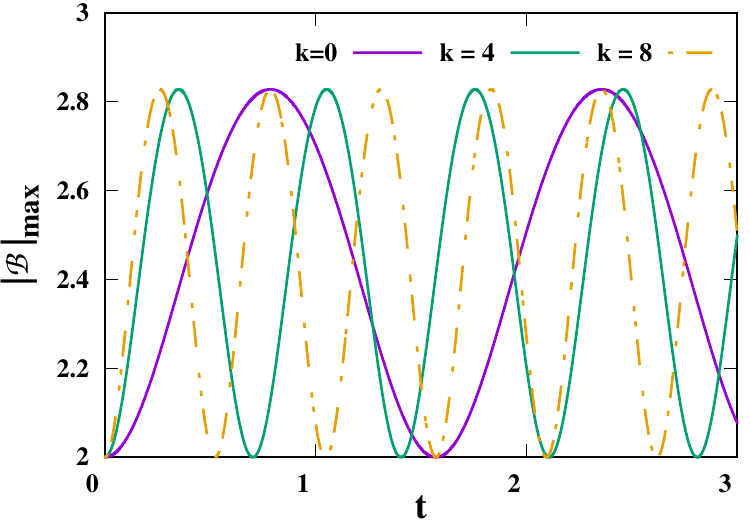}
\caption{(Color online.) Dynamics of $|\mathcal{B}|_{\max}$ (ordinate) with time, $t$ (abscissa). The initial state  is taken to be  the $k$th Fock 
state in the CV part. We choose different values of $k=0,4,$ and $8$, which illustrate  different oscillation periods of $|\mathcal{B}|_{\max}$. Both axes are dimensionless. }
\label{fig:fixed_k} 
\end{figure}

If the electromagnetic field is initally in the $k{\text{th}}$ Fock basis, i.e.,
 \begin{eqnarray}
 C_{n} = \delta_{n,k},   
 \label{eq:kfock}
 \end{eqnarray}
the time evolved hybrid state reads as
 \begin{eqnarray}
 \nonumber
    \ket{\psi(t)} &=& \cos(\lambda t \sqrt{k+1})\ket{k}\ket{e} \\ &-& i\sin(\lambda t \sqrt{k+1})\ket{k+1}\ket{g},
    \label{eq:fixed_k}
 \end{eqnarray}
 which leads to the maximal violation at any instant of time as
 \begin{eqnarray}
     | \mathcal{B} |_{\max} = 2\sqrt{1+\sin^2\big({2 \lambda  \sqrt{1+k} t}\big)}.
     \label{eq:BVfock}
 \end{eqnarray}
\ph{It is obtained  by calculating the two largest eigenvalues of \(T^{\dagger}T\) of \(|\psi(t)\rangle\) in Eq. (\ref{eq:fixed_k}).}
We find that the state becomes entangled for all $k$ and the entire range of $t$ values except for a few distinct values, $|\mathcal{B}|_{\max}>2$ as also shown in Fig. \ref{fig:fixed_k}. Depending on $k$, only the period of oscillation changes as seen from Eq. (\ref{eq:BVfock}).
\ph{Specifically, at times $t=\frac{n\pi}{2\lambda\sqrt{1+k}}$ where $n$ is any integer, the state becomes unentangled having $|\mathcal{B}|_{\max}=2$ while $|\mathcal{B}|_{\max}=2\sqrt{2}$ at $t=\frac{\frac{\pi}{2}+2n\pi}{2\lambda\sqrt{1+k}}$ with period of oscillation of $|\mathcal{B}|_{\max}$ given by $\frac{\pi}{2\lambda\sqrt{1+k}}$. Note that the corresponding example signifies an effective two-qubit system where the continuous variable part also reduces to two dimensions of the system too. Since all pure two-qubit entangled states must violate CHSH Bell inequality \cite{Gisin1991,rt1} which implies that when $|\mathcal{B}|_{\max}=2$, the state must be separable. }

 


 \paragraph{Initial field prepared in SMSV state.}
 Let us move to the scenario where the initial CV system is in a single-mode squeezed vacuum (SMSV) state having the form
 \begin{eqnarray}
 \ket{\psi(0)}_{\text{field}} &=& {S}(\xi)\ket{0} = \ket{\xi} = \sum_{n=0}^\infty C_n \ket{n},
 \end{eqnarray}
 \begin{eqnarray}
 \text{with  }    C_{n} = \begin{cases}
     
     (-1)^{n/2}\frac{\sqrt{n!}}{2^{\frac{n}{2}}\frac{n}{2}!} \frac{(e^{i\theta} \tanh(r))^{\frac{n}{2}}}{\sqrt{\cosh(r)}} & \text{if n is even ,}\\\nonumber
     0 & \text{if n is odd,}
     \end{cases}\\
 \end{eqnarray}
 \begin{eqnarray}
\text{where }     S(\xi) = \exp\bigg[\frac{1}{2}(\xi^* a^2 - \xi a^{\dagger 2})\bigg], \hspace{0.6cm} \xi = r e^{i \theta}.
 \end{eqnarray}
Following the prescription, described in Sec. \ref{sec:maximization_bell}, we find the maximal Bell violation for each $q$ denoted by $|\mathcal{B}|_{\max}^q$ where optimizations over $\bm{\theta}$, $\bm{\Phi}$ are already performed.
For $q=0$, the correlation matrix takes the form as $T =    \begin{pmatrix}
0 & \epsilon & 0 \\
-\epsilon & 0 & 0 \\
0 & 0 & -1
\end{pmatrix}$ and the corresponding $|\mathcal{B}|_{\max}^0 = 2\sqrt{1+\epsilon^2}$ with the directions being derived as $\bm{\theta} = (\pi,\frac{\pi}{2},\tan^{-1}\epsilon,-\tan^{-1}\epsilon)$ and $\bm{\Phi} = (\phi, 0, \frac{\pi}{2}, \frac{\pi}{2})$, \(0\leq \phi \leq 2 \pi\) which implies that the maximization is independent of \(\phi\).
In case of $q=1$, $T = \begin{pmatrix}
\kappa_1 & \kappa_2 & 0 \\
\kappa_2 & -\kappa_1 & 0 \\
0 & 0 & \kappa_3
\end{pmatrix}$. If $\kappa_3^2 > \kappa_1^2 + \kappa_2^2$, we obtain $|\mathcal{B}|_{\max}^1 = 2\sqrt{\kappa_3^2 + (\kappa_1^2 + \kappa_2^2)}$ and optimizing directions are $\bm{\theta} = (0, \frac{\pi}{2}, \tan^{-1}\sqrt{\kappa_1^2+\kappa_2^2}, -\tan^{-1}\sqrt{\kappa_1^2+\kappa_2^2})$, $\bm{\Phi} = (\phi, \tan^{-1}(-\frac{\kappa_1}{\kappa_2}), \frac{\pi}{2}, \frac{\pi}{2})$.
Finally, for $\kappa_3^2 < \kappa_1^2 + \kappa_2^2$, the maximal Bell violation reads as $|\mathcal{B}|_{\max}^1 = 2\sqrt{2}\sqrt{\kappa_1^2 + \kappa_2^2}$ along with the optimal measurement directions being calculated as $\bm{\theta} = (\frac{\pi}{2}, \frac{\pi}{2}, \frac{\pi}{2}, \frac{\pi}{2})$ and $\bm{\Phi} = (\tan^{-1}\frac{\kappa_2}{\kappa_1}, \tan^{-1}(-\frac{\kappa_1}{\kappa_2}), \frac{\pi}{4}, \frac{7\pi}{4})$. Thus depending on the above conditions, we get $|\mathcal{B}|_{\max} = \max \{ |\mathcal{B}|^0_{\max}, |\mathcal{B}|^1_{\max}\}$ where the range of $q$ from which the maximal violation obtained, is confirmed via numerical investigation.

\begin{figure}
\includegraphics[width=1.0\linewidth]{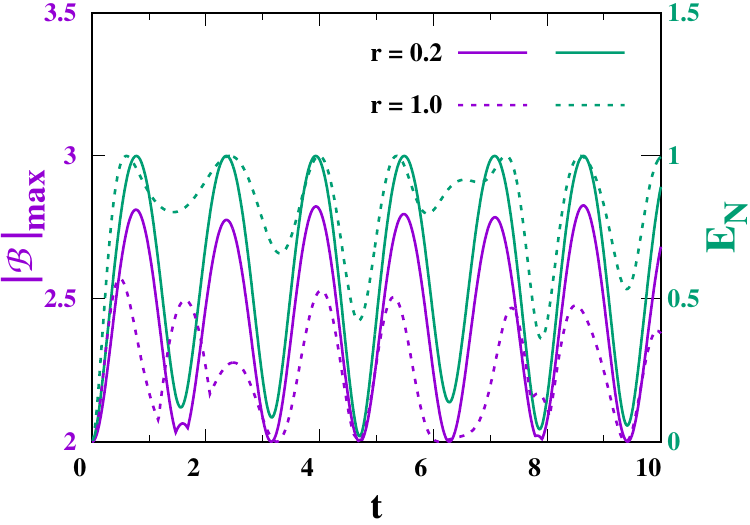}
\caption{(Color online.)  Dynamics of maximal violation of Bell inequality,  $|\mathcal{B}|_{\max}$ (left ordinate\ph{, purple color})  with time $t$ (abscissa) of the time evolved hybrid state, where the initial state in the CV sector is a single mode squeezed vacuum (SMSV) state with the atom initially being in the excited state. 
We consider two different values of squeezing strength, $r = 0.2\ph{\text{ (solid line)}},\, \mbox{and} \,  1.0\ph{\text{ (dashed line)}}$. We plot the entanglement content, $E_N$ (right ordinate\ph{, teal color}), quantified by the von Neumann entropy of the local density matrix of the time evolved state against time which connects \(E_N\) with $|\mathcal{B}|_{\max}$. \ph{Note that the maximum of the left ordinate can go upto \(2\sqrt{2}\) while the right ordinate  is bounded above by \(1\). } Both axes are dimensionless. }
\label{fig:smsv} 
\end{figure}

In Fig. \ref{fig:smsv}, we show the time dynamics of the maximal Bell expression and show that during the entire evolution $|\mathcal{B}|_{\max} \geq 2$. We also compare the dynamical violation with the entanglement content of the state. Since the time evolved state is pure, the entanglement can simply be quantified by the entanglement entropy, i.e., the von Neumann entropy of the reduced density matrix given by $E_N = \mathcal{S}(\rho_d(t)) = -\text{Tr} [\rho_d(t)\log_2\rho_d(t)]$ with $\rho_d(t)$ being the reduced density matrix of the qubit subsystem of the time evolved state.
Moreover, with the increase of squeezing strength, the maximal value, $|\mathcal{B}|_{\max}$, for a fixed time interval depicts an overall declining feature, although the fall is not always monotonic. 
Interestingly, if we keep on increasing $r$, we observe that after a certain critical value $r=r_c$ of the field, the maximal violation for the hybrid system completely vanishes. 


\begin{figure}
\includegraphics[width=1.0\linewidth]{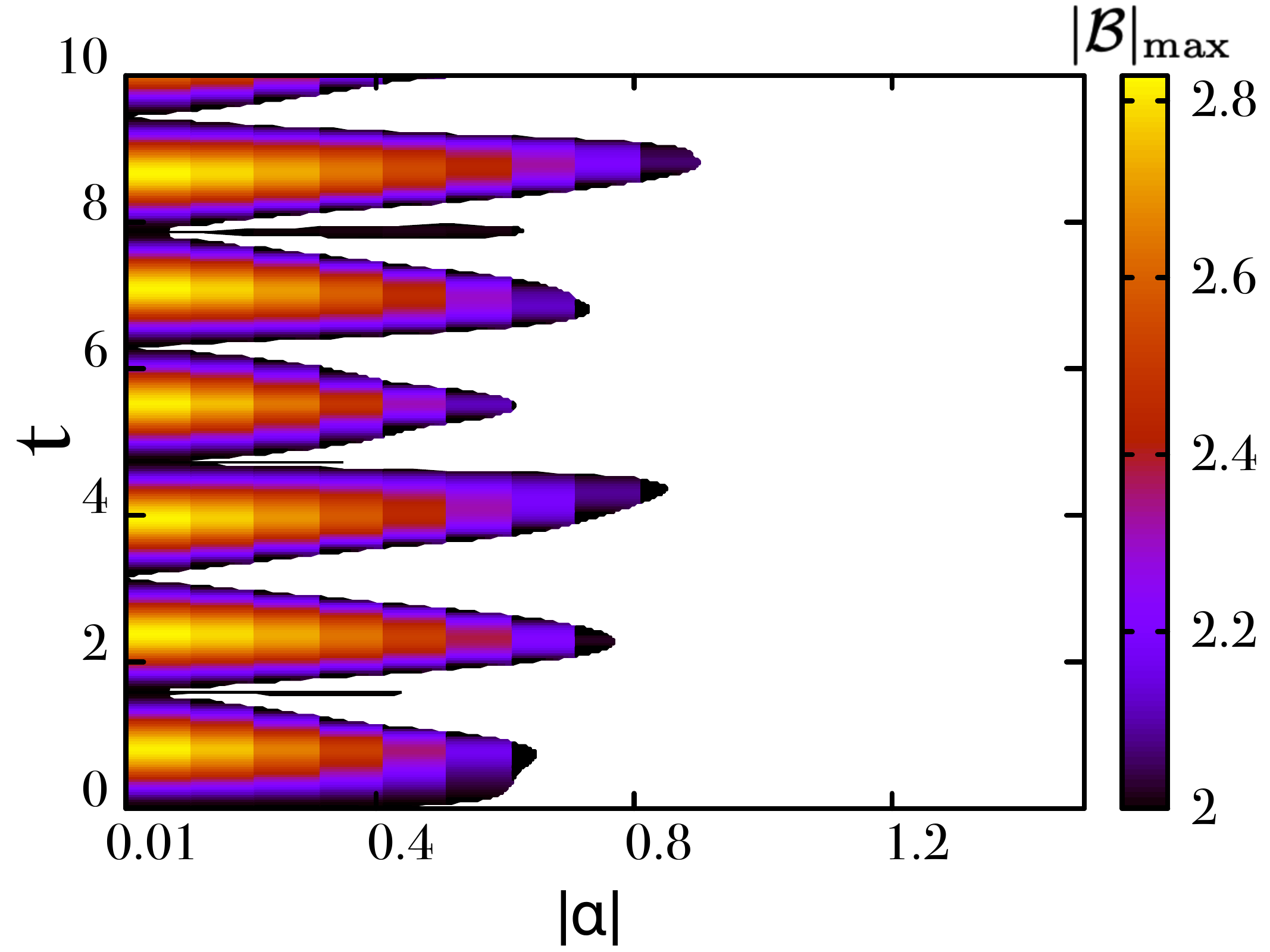}
\caption{(Color online.) A heat map plot of  $|\mathcal{B}|_{\max}$ in the $(|\alpha|, t)$- (horizontal-vertical axis) plane where the initial CV system is  taken as a coherent state which evolves under JC Hamiltonian. The uncolored (white) regions denote the non-violating zones.   Both axes are dimensionless.}
\label{fig:diff_alpha} 
\end{figure}

 \paragraph{Coherent state as initial field state.}
 Let us choose the displaced vacuum in the Fock basis, 
 \begin{equation}
\ket{\psi(0)}_{\text{field}} = D(\alpha)\ket{0} = \ket{\alpha}
\label{eq::coherent}
 \end{equation}
 where $\ket{\alpha} = \sum_{n=0}^\infty C_n \ket{n}$, with $C_n = \frac{e^{-|\alpha^{2}|} \alpha^{n}}{\sqrt{n!}}$, $\alpha = |\alpha|e^{i\theta}$ as the initial field state. In this case, the number of optimizing parameters cannot be reduced like in the previous scenario. 
Therefore, we perform the optimization using numerical techniques where we find the maximal violation by directly finding the eigenvalues of the correlation matrix using standard numerical eigensolvers.  
Moreover, we scan the space of $q \in \mathbb{N}$ and find the value of $q$ that optimizes the Bell expression to be between $0$ to $5$.
For a particular time interval, optimizing over the measurement settings, $\bm{\theta}, \bm{\Phi}, q$, and by varying the displacement strength $|\alpha|$ of the coherent state, $|\mathcal{B}_{\max}|$, shows an oscillatory behaviour (see Fig. \ref{fig:diff_alpha}).

\begin{figure}
\includegraphics[width=1.0\linewidth]{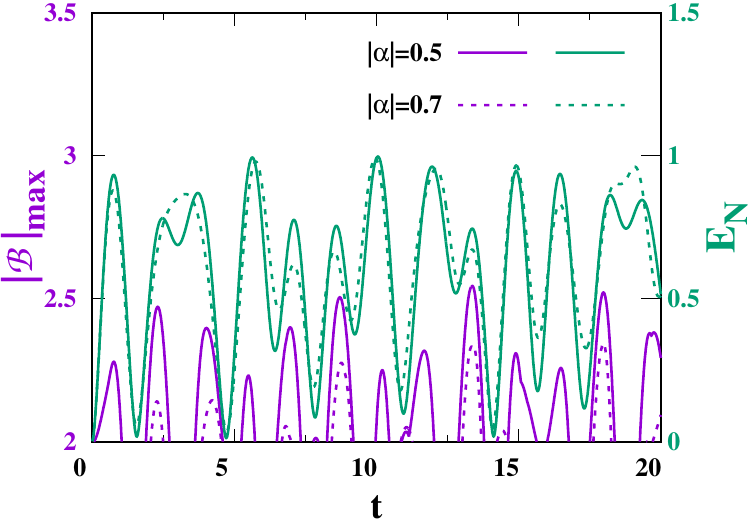}
\caption{(Color online.) Connecting  entanglement (right ordinate) with the maximal violation of Bell inequality (left ordinate) for initial coherent states with two prototypical intensities: $|\alpha|=0.5$ and $|\alpha|=0.7$. A similar analysis for an initial SMSV state is demonstrated in Fig. \ref{fig:smsv}. Both axes are dimensionless.}
\label{fig:ent_vs_bell_a0.5} 
\end{figure}

Despite being entangled at certain times, it does not violate Bell inequality at those times due to the fact that the pseudospin operators do not form a basis for all dichotomic operators in space of single mode  continuous variable state in the first party. 
For a particular time interval, the total time span at which we are detecting the Bell nonlocality through hybrid spin formalism decreases with the increase of displacement of the coherent state since it fails to completely capture the full two-party correlation, as shown in Fig. \ref{fig:ent_vs_bell_a0.5}. Furthermore, we notice that there is a critical value of $|\alpha|$ beyond which the state does not violate the  Bell inequality (see Fig. \ref{fig:diff_alpha}).



\ph{\textbf{Remark:} Note that, in case of SMSV (Fig.~\ref{fig:smsv}), entanglement and nonlocality are out of sync in the early times, especially for $r=1.0$. Along with that, there exists a non-oscillatory pattern of nonlocality in certain parameter regimes in the case of the initial state as a coherent state. 
These observations are possibly due to the suboptimality of the pseudospin measurements on the CV subsystem of the hybrid system.
Although the set of parameterized pseudospin operators constitutes valid measurements in the CV sector, they do not exhaust the set of all possible dichotomic measurements on CV systems. Therefore, when we maximize the measurement choices in the subset of pseudospin measurements, the obtained value remains suboptimal with respect to all possible dichotomic measurements in the CV sector. So, the obtained value of violation of hybrid nonlocality lower bounds the theoretically achievable maximum. The intuition about the oscillatory behavior of nonlocality and its synchronization with entanglement holds true, only when for  nonlocality, the maximal violation is achieved. Since the best pseudospin measurements can be non-optimal  and,  consequently, the obtained value of the violation is irregular and out of sync with entanglement. Here we also want to stress that as per literature, and our own understanding, obtaining the globally optimal dichotomic measurements in the CV sector is  realistically impossible to do.}

\subsubsection{Classically correlated states as inputs}
\begin{figure}
\includegraphics[width=1.0\linewidth]{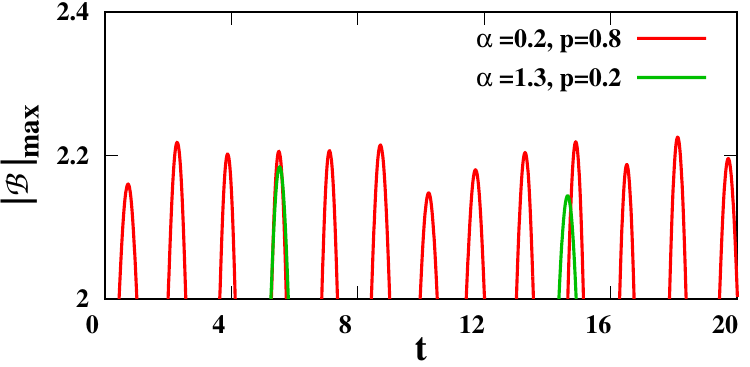}
\caption{(Color online.)
For initial classically correlated states, \(|\mathcal{B}|_{\max}\) (vertical axis) vs \(t\) (horizontal axis). We observe that only for low $\alpha$ - high $p$, and high $\alpha$ - low $p$ values,  there is a possibility for obtaining a finite violation of local realism. Both axes are dimensionless.}
\label{fig:cl_cor} 
\end{figure}
Instead of product states, let us allow the initial hybrid state to share some classical correlation which can be prepared via local operation and classical communication (LOCC). For example, we take 
\begin{equation}
    \rho(0) = p \ket{e,\alpha}\bra{e,\alpha} + (1-p)\ket{g,-\alpha}\bra{g,-\alpha}.
\end{equation}
Here we take three different values of $p = 0.2,0.5$ and $0.8$. For $p=0.2$, the state is a convex mixture of two different separable states which is close to a pure state $\ket{g,-\alpha}$. For $p=0.5$, it is an equal mixture and $p=0.8$, the classically correlated state has less distance with a pure state, $\ket{e,\alpha}$. After the evolution of the initial state governed by Jaynes-Cummings Hamiltonian, we  numerically compute the maximal Bell-CHSH expression at each time using Horodecki's criterion \cite{horodecki1995}.

Depending on the displacement strength $|\alpha|$ of the initial  CV system, we have different time dynamics of this maximal Bell expression. As shown in Fig. \ref{fig:cl_cor}, with low displacement $|\alpha|=0.2$, and high $p=0.8$, we get a $|\mathcal{B}|_{\max}$ which crosses LHV bound $2$ during the dynamics. For $p=0.5$ the same $|\alpha|$ value does not lead to a violation.  Reversely, with high displacement $|\alpha|=1.3$, $|\mathcal{B}|_{\max} > 2 $ for $p=0.2$ while with $p=0.5$ and $0.8$, $|\mathcal{B}|_{\max} < 2$. In summary, violation of local realism with spin-pseudospin measurement is achieved for low $p$ and high $|\alpha|$ or with high $p$ and low $|\alpha|$. This is a prototypical example to show the efficacy of our method to detect the violation of local realism with mixed hybrid state having classical correlation.


\subsection{Dynamics of hybrid nonlocality for initially entangled states}
\label{sec:initialent}

We now change the gear by choosing the initial state as an entangled state, instead of the initial separable states studied before. 
Let us consider the cat state \cite{Gerry},
\begin{eqnarray}
    \ket{\psi(0)} = a_1\ket{\alpha}\ket{e} + a_2\ket{-\alpha}\ket{g},
    \label{eq:catgen}
\end{eqnarray}
as the initial state having bipartite entanglement. In Schrodinger picture, the evolved state reads as $\ket{\psi(t)} = e^{-i H_0 t} U_I(t) \ket{\psi(0)} = \ket{\psi_g(t)}\ket{g} + \ket{\psi_e(t)}\ket{e}$,
\begin{eqnarray}
\text{where  }\nonumber
    \ket{\psi_g(t)} &=& \big[ \sum_{n=0}^{\infty} e^{-i\omega(n-1/2)t} \{-i C_{n-1} p_1 \sin{(\lambda t \sqrt{n})} \\ \nonumber &+& (-1)^n p_2 C_{n} \cos{(\lambda t \sqrt{n})}\}\ket{n}\big], \\ \nonumber
\text{and  }    \ket{\psi_e(t)} &=& \big[ \sum_{n=0}^{\infty} e^{-i\omega(n+1/2)t} \{C_{n} p_1 \cos{(\lambda t \sqrt{n+1})} \\ \nonumber &-& i(-1)^{n+1} p_2 C_{n+1} \sin{(\lambda t \sqrt{n+1})}\}\ket{n}\big],\\
\label{eq:psig_psie}
\end{eqnarray}
\begin{eqnarray}
 \text{with  }    C_n = \frac{e^{-|\alpha^{2}|} \alpha^{n}}{\sqrt{n!}}.
\end{eqnarray}
\begin{figure}
\includegraphics[width=1.0\linewidth]{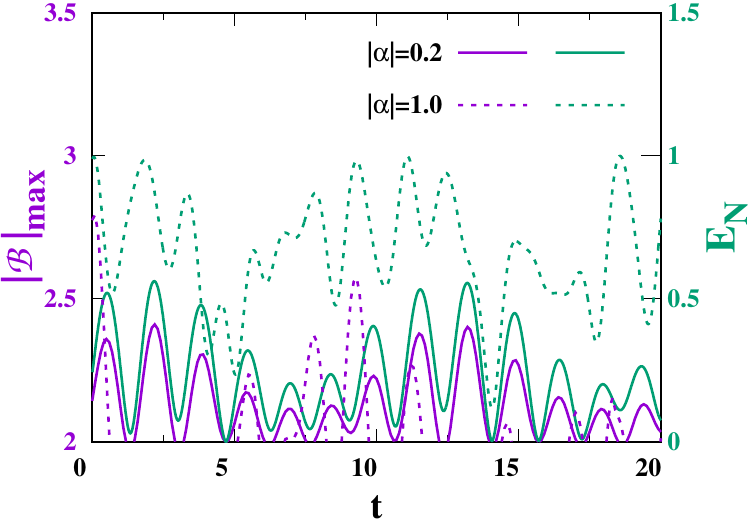}
\caption{(Color online.)  Dynamics of the maximal violation of local realism $|\mathcal{B}|_{\max}$ (left ordinate)  with time $t$ (abscissa) of the hybrid  cat state for different values of  initial displacement $|\alpha|= 0.2, \, \mbox{and}, \, 1.0$. For comparison, entanglement, $E_N$ (right ordinate), of the same time evolved cat state is shown. \ph{Rest of the specifications are similar to Fig.~\ref{fig:smsv}.} Both axes are dimensionless.}
\label{fig:cat_bell_violation} 
\end{figure}Note that irrespective of the value of frequency $\omega$ of JC Hamiltonian, we get the same $|\mathcal{B}|_{\max}$ at all times with $a_1=a_2=\frac{1}{\sqrt{2}}$ which is due to 
the structure of the time evolved state. Firstly, the qubit subsystem can be written as 
\begin{eqnarray}
\nonumber
    \rho_q(t) &=& |\psi_g(t)|^2 \ket{g}\bra{g} + |\psi_e(t)|^2 \ket{e}\bra{e} \\  &+& \langle \psi_e(t)|\psi_g(t)\rangle \ket{g}\bra{e} + \langle \psi_g(t)|\psi_e(t)\rangle \ket{e}\bra{g}, \nonumber \\
\end{eqnarray}
where
\begin{eqnarray}
\nonumber
    \langle \psi_g(t)|\psi_e(t)\rangle &=& \sum_{n=0}^\infty e^{-i \omega t} \{i C^*_{n-1} p_1 \sin{(\lambda t \sqrt{n})} \\ \nonumber &+& (-1)^n p_2 C^*_{n} \cos{(\lambda t \sqrt{n})}\} \\ \nonumber
    &\times& \{C_{n} p_1 \cos{(\lambda t \sqrt{n+1})} \\ \nonumber &-& i(-1)^{n+1} p_2 C_{n+1} \sin{(\lambda t \sqrt{n+1})}\}.\\
\end{eqnarray}
When we derive the eigenvalues of $\rho_d$, we find that the phase factors with $\omega$ cancel out, which implies that the entanglement of the cat state is independent of the resonant frequency. Therefore, the entanglement of $\ket{\psi(t)}$ reduces to $E_N = \mathcal{S}(\rho_d(t))$.
Fig. \ref{fig:cat_bell_violation} again confirms that the trends of the overall dynamics of entanglement and the maximal violation of Bell inequality qualitatively match.
In this situation also, we find that with increasing $|\alpha|$, the capability of capturing nonlocal correlation decreases by hybrid spin formalism.
\begin{figure}
\includegraphics[width=1.0\linewidth]{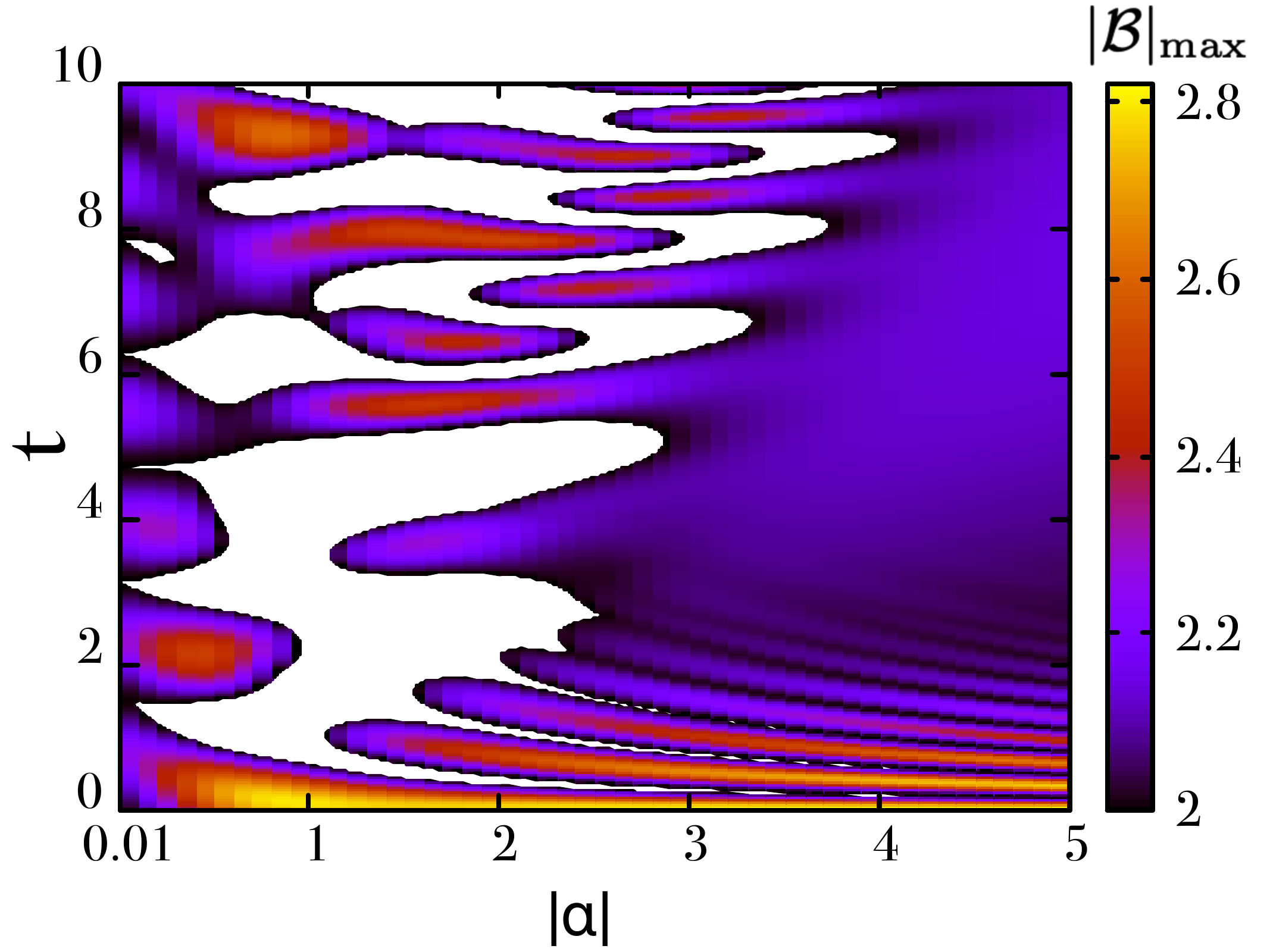}
\caption{(Color online.) Map plot of the maximal violation of Bell inequality\ph{, $|\mathcal{B}|_{\max}$} in the $(|\alpha|, t)$-plane for an initial cat state. The uncoloured (white) regions represent the non-violating river. By comparing with Fig. \ref{fig:diff_alpha}, we notice that for initial product state, no violation is observed for high \(|\alpha|\) values \(\forall t\) while in case of initial entangled state, the dynamical state always violates local realism, irrespective of \(|\alpha|\) and \(t\) values.     Both axes are dimensionless.}
\label{fig:t=fixed_cat_bell_violation} 
\end{figure}
Moreover, from Fig. \ref{fig:t=fixed_cat_bell_violation}, it is evident that at time $t=0$ with increasing $|\alpha|$, the maximal violation of Bell inequality rapidly increases, when $|\alpha|$ is small and then after showing a small dip, it saturates to $2\sqrt{2}$ for a certain range of displacement strength.
Interestingly, from Fig. \ref{fig:t=fixed_cat_bell_violation}, we observe that for low values of $|\alpha|$, which corresponds to lower values of initial entanglement, a river of non-violating regime in the $(|\alpha|,t)$-plane emerges where no violation of Bell-CHSH inequality is found. However, when the initial entanglement is high indicated by higher $|\alpha|$ values, we obtain violation for all times.


\section{Based On Parity Operator, Connecting Hybrid Wigner non-classicality with  violation of local realism}
\label{sec:Wigner}
Investigation of the connection between negativity of the Wigner function dates back to the time of J. S. Bell who argued that Gaussian states would not violate a Bell inequality since they possess a Gaussian (positive) Wigner function \cite{Bell1964}. However, Banaszek et. al.   \cite{Banaszek1998} conclusively showed that two-mode Gaussian states can violate Bell inequalities, thereby demonstrating a weak connection between the sign of the Wigner function of two-mode CV state and local realism. We intend to explore  the possibility of establishing a connection between the hybrid versions of these quantities for bipartite states constituting of both discrete and continuous variable systems.

By considering all the initial product states and cat states, we evaluate $ \mathcal{V}_n$ defined in Eq.~(\ref{eq:negvol}) and compare it with maximal violation of Bell inequality. The dynamical behaviour is similar for all initial states. For prototypical demonstration, we chose the initial state to be the cat state. Notice that unlike product states, this cat state violates local realism even at the initial time.

 By substituting $a_1 = a_2 = \frac{1}{\sqrt{2}}$ in Eq. \eqref{eq:catgen}, the time evolved cat state can be written as
\begin{eqnarray}
    \ket{\psi(t)}_{\mathcal{C}} = \sum_{m=0}^\infty C_m^g(t) \ket{m}\ket{g} + \sum_{m=0}^\infty C_m^e(t) \ket{m}\ket{e}.
\end{eqnarray}
The dynamical hybrid Wigner function for the time-evolved cat state can be written from Eq. \eqref{eq:hybridwigner} as 
\begin{eqnarray}
    W(\theta, \phi, \beta) = \bra{\psi(t)}\hat{\Delta}_c(\beta)\otimes \hat{\Delta}_d(\theta, \phi)\ket{\psi(t)},  
\end{eqnarray}
where  $\hat{\Delta}_c(\beta)$ and $\hat{\Delta}_d(\theta, \phi)$ denote the kernel operators corresponding to the continuous and qubit systems respectively. It reduces to
\begin{eqnarray}
\nonumber
    W(\theta,\phi,\beta) &=& \sum_{n,m=0}^\infty {C_n^g}^* (t) C_m^g(t) \bra{n}\hat{\Delta}_c(\beta)\ket{m} \\ \nonumber &\times& \Big(\bra{g}\hat{\Delta}_d(\theta,\phi)\ket{g} + \bra{g}\hat{\Delta}_d(\theta,\phi)\ket{e} \\ &+& \bra{e}\hat{\Delta}_d(\theta,\phi)\ket{g} + \bra{e}\hat{\Delta}_d(\theta,\phi)\ket{e}\Big).
\end{eqnarray}
By evaluating,
\begin{eqnarray}
\nonumber
    \bra{n}\hat{\Delta}_c\ket{m} &=& \frac{2}{\pi}\sum_{n'=0}^\infty (-1)^{n'}\bra{n}{\hat{D}(\beta)}\ket{n'}\bra{n'}\hat{D}^\dagger(\beta)\ket{m} \\
    &=& \frac{2}{\pi}F(n,m,\beta),
\end{eqnarray}
where $F(n,m,\beta) = \sum\limits_{n'=0}^\infty (-1)^{n'}\bra{n}{\hat{D}(\beta)}\ket{n'}\bra{n'}\hat{D}^\dagger(\beta)\ket{m}$. The elements of $\hat{D}(\beta)$ are given by
\begin{eqnarray}
\nonumber
    \bra{k}\hat{D}(\beta)\ket{l} &=& \bra{k}\ket{\beta,l} \\ \nonumber
    &=& \begin{cases}
        \bra{0}\ket{\beta}\sqrt{\frac{l!}{k!}}\beta^{k-l}L_l^{k-l}(|\beta|^2), & k>l\\
        \bra{0}\ket{\beta}\sqrt{\frac{k!}{l!}}{(-\beta^*)}^{l-k}L_k^{l-k}(|\beta|^2), & l>k\\
    \end{cases}\\
\end{eqnarray}
with $\bra{\beta}\ket{\alpha} = e^{-\frac{1}{2}(|\alpha|^2+|\beta|^2+\alpha {\beta^*})}$ and $L_m^n(x)$ is the Laguerre polynomial $L_m^n(x) = \sum_{i=0}^m \begin{pmatrix}
    m+n \\
    m-n
\end{pmatrix} \frac{(-x)^i}{i!}$.
Finally, we obtain 
\begin{eqnarray}
\nonumber 
&&W(\theta, \phi, \beta) \\ \nonumber 
&&= \frac{1}{\pi}\sum_{n,m=0}^\infty \{(1+\sqrt{3}\cos{2\theta}){C_n^g}^*(t){C_m^g}(t)\\ \nonumber 
&&+ (1-\sqrt{3}\cos{2\theta}){C_n^e}^*(t){C_m^e}(t) \\ \nonumber
&&+(\sqrt{3}\sin{2\theta}e^{i2\phi}){C_n^e}^*(t){C_m^g}(t) \\ \nonumber
&&+(\sqrt{3}\sin{2\theta}e^{-i2\phi}){C_n^g}^*(t){C_m^e}(t)\}F(n,m,\beta).
\end{eqnarray}
By integrating the negative part of the Wigner function over the Haar measure $d\Omega$ as defined in Eq. \eqref{eq:measure}, we find the hybrid Wigner negativity, $\mathcal{V}_n$ in Eq.~(\ref{eq:negvol}).

\begin{figure}
\includegraphics[width=1.0\linewidth]{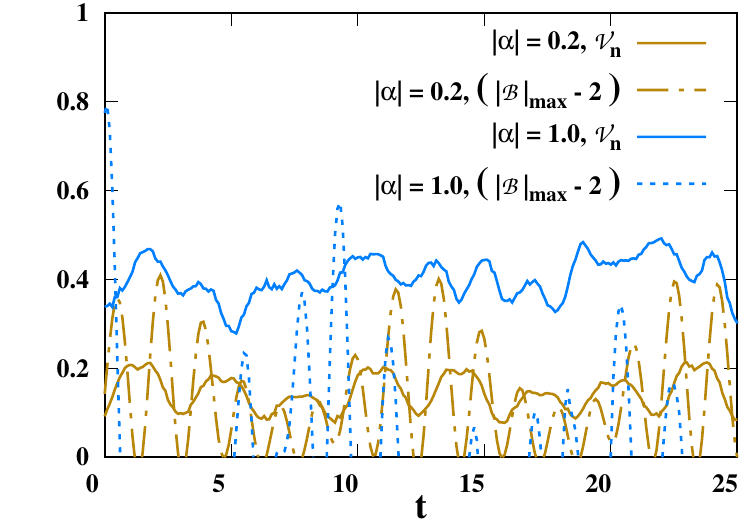}
\caption{(Color online.) Connection between the maximal violation of local realism, $|\mathcal{B}|_{\max}-2$ (vertical axis) (dashed lines), and negative Wigner volume $\mathcal{V}_n$ (vertical axis) indicated by solid lines, with time, $t$ (horizontal axis) by taking the initial CV system to be in a cat state. Both axes are dimensionless.}
\label{fig:comparison} 
\end{figure}
Our investigation \ph{(see appendix \ref{sec:app})} clearly reveals that there is no one to one connection between the maximal violation of Bell inequalities and the negative Wigner volume in hybrid systems   (see Fig.\ref{fig:comparison}) as also shown for CV systems \cite{Banaszek1998}. 


\section{Effect of Disorder in atom-photon interactions}
\label{sec:imperfection}
Till now, we have considered an idealized situation in which the quantum states are free from any noise and the time dynamics do not have any imperfection or disorder. However, in a realistic situation, imperfections are ubiquitous at all stages of quantum technology. This makes the investigation of imperfections and their corresponding robustness analysis absolutely 
crucial. In our work, we 
consider the situation where the coupling strength of the atom-photon interaction has some impurity. Such a situation might arise when the cavity is not fabricated perfectly resulting in varied imperfections like the field inside the cavity  not being uniform or the disorder being present in the coupling strength. \ph{Specifically, stochastic fluctuations in the interaction strength may be inherited from the source of a single mode coherent field coupled to a cavity or due to the variations in the mechanisms involved in the production of Rydberg atoms, which may arise from instabilities in the production of atomic vapor \cite{quang1991,joshi1993,joshi1994,ahana}. The disorder in the interaction strength can be modelled by $\lambda(t) = \lambda_0(t)e^{-i\phi(t)}$, where $\lambda_0(t)$ and the phase $\phi(t)$ are stochastic variables \cite{joshi1993,joshi1994}.
However, in our work, we assume that -- (i) there is no fluctuation in phase, i.e., $\phi(t)=0$ $\forall t$ and (ii) the amplitude $\lambda(t)$ has some fluctuations which can be modelled as 
\begin{eqnarray}
    \nonumber \lambda(t) = \lambda \text{   with probability distribution }P(\lambda), \text{   for }0\leq t \leq t_O,\\
\end{eqnarray}
where $t_O$ denotes the time of observation of a single run of the experiment. We assume that $\lambda(t)$ remains fixed for the entire time of observation of a particular run of the experiment.}
We now examine whether the maximal violation of Bell inequality is robust against these imperfections.




\ph{The value of maximal violation of Bell inequality, i.e., $|\mathcal{B}_{\max}|$ at different times 
in case of }such disordered systems are obtained by averaging over various realizations of the disorder,
 namely quenched averaging. It is performed when the equilibration time of the disorder which  in the Jaynes-Cummings interaction strength, $\lambda$ is much longer compared to the observation time scales of the physical properties of the system. 
See previous work \cite{ahana} in this respect. 
Therefore, one may consider the value of the impurities in the system to be effectively fixed during the dynamics of the system, thereby making it possible to carry out an averaging of the quantity of interest, $\mathcal{Q}$ \ph{(which is the violation of Bell inequality, $|\mathcal{B}|$ in our work)}, over the distribution of different values of the disordered parameters. Such a disordered system is realizable  in current experimental setups using systems like cold atoms and trapped ions \cite{Aspect08, deMartini2008, Lewenstein2007, Aspect2009}. For randomly chosen  interactions $\lambda$ from a probability distribution $P(\lambda)$, the quenched-averaged $\mathcal{Q}$, denoted by \(\langle \mathcal{Q}\rangle\),  at every time instant is given by
\begin{equation}
\langle \mathcal{Q}\rangle = \int\mathcal{Q}(\lambda)P(\lambda)d\lambda.
\label{eq:dis}
\end{equation}
Note that $P(x) = \delta(x - \lambda)$ corresponds to the disorder free case, which were discussed in the previous sections. To capture the effect of disorder, various forms of $P(\lambda)$ are chosen in the literature. The typical choices include uniform distribution, Gaussian distribution, Cauchy-Lorentz distribution \cite{sachdev_2011,Nishimori_2001,spin_glass_1986,disorder_1,BKC1996} .
\textcolor{black}{For our analysis, we will restrict our attention to the Gaussian distribution of interaction strengths. Let us explain our choice briefly. When various fluctuating factors determining the interaction strength meet, the resulting distribution is typically Gaussian. This is a crude way of stating the central limit theorem \cite{Kardar2007}, which ensures that the sum over many random variables with arbitrary distributions has a simple Gaussian probability distribution function.}
In our work, we choose the values of $\lambda$ from a Gaussian distribution with mean $\bar{\lambda}$ and standard deviation $\sigma_{\lambda}$, which can be called the strength of the disorder. Unlike entanglement measures \cite{horodecki-ent}, two different situations \ph{-- (a) oracle measurement strategy, and (b) realistic measurement strategy} can be considered in this set up since one requires measurement outcomes to obtain Bell statistics.

\subsection{Oracle measurement strategy}
Depending on the value of the interaction strength in each realization $\lambda$, the state evolved through Jaynes-Cummings Hamiltonian and its time dependent correlation matrix can be written respectively as 
\begin{eqnarray}
      \mathit{T}^{ij}_{\lambda}(t)&=& \text{Tr}(\rho_{ \lambda}(t) S_{i}^q \otimes\sigma_{j}),\nonumber\\
 \text{and } \rho_{\lambda}(t) &=& U_{\lambda}(t)\rho_{in}U^\dagger_{\lambda}(t).
\label{eq:tmatrixdis}
\end{eqnarray}
Let us now suppose that the experimentalist has the ability to tune its measurement directions to reach the maximal Bell violation at any given time $t$ depending on the particular realization $\lambda$. 
Hence, for every realization $\lambda$ and at any given time instance $t$, we can apply the optimization strategy as laid out in Sec. \ref{sec:max} and we obtain
\begin{equation}
     \mathcal{Q}^O(\lambda) =|\mathcal{B}_{\rho_{\lambda}}|_{\max} = 
      2 \sqrt{\Lambda_1 + \Lambda_2},
\label{eq:QAoracle}
\end{equation}
where $\Lambda_1$ and $\Lambda_2$ denote the two largest eigenvalues of $T^{\dagger}_{\lambda}T_\lambda$, and $O$ in the superscript denotes the oracle measurement strategy. 
Note that we drop the time index since it is fixed to a given point $t$. 
Nevertheless, for the disordered case, when the measurement settings can be chosen depending on a given realization $\lambda$, we are now equipped to calculate the quenched averaged maximal violation by using Eq. \eqref{eq:dis} as
\begin{equation}
\langle \mathcal{Q}\rangle^{O} = \int\mathcal{Q}^{O}(\lambda)P(\lambda)d\lambda.    
\end{equation}
In this work, we consider the exemplary case where the disorder follows a Gaussian distribution,
\begin{eqnarray}
 P(\lambda) = \frac{1}{\sigma_\lambda \sqrt{2\pi}}\exp \frac{(\lambda - \bar{\lambda})^2}{2 \sigma_{\lambda}^2},   
\end{eqnarray}
where  $\bar{\lambda}$ is the mean and the corresponding standard deviation is $\sigma_{\lambda}$.
Here $\sigma_{\lambda}=0$ corresponds to the ordered case discussed in previous sections. We now present some examples to demonstrate how the disorder effects the violation of the Bell-CHSH inequality. 

The first example we choose is when the initial state is a product state where the state of the field mode is in the $k^{\text{th}}$ Fock basis (see Eq. \eqref{eq:kfock}) and the atom is in the excited state. The maximal violation of local realism for any realization $\lambda$ at any time slice $t$ is given in Eq. \eqref{eq:BVfock}.  Therefore we have, $Q(\lambda) = 2\sqrt{1+\sin^2\big({2 \lambda  \sqrt{1+k} t}\big)}$. The quenched average violation can be expressed as
\begin{eqnarray}
&& \langle \mathcal{Q}\rangle^{O} = \frac{2}{\sigma_\lambda \sqrt{2\pi}} \times \nonumber \\
&& \int_{-\infty}^{\infty} \exp \frac{(\lambda - \bar{\lambda})^2}{2 \sigma_{\lambda}^2} \sqrt{1+\sin^2\big({2 \lambda  \sqrt{1+k} t}\big)} d\lambda.
\label{eq:QAk}
\end{eqnarray}
Since there is no closed form expression of this integral, we resort to standard numerical techniques to perform this integration.
\begin{figure}
\includegraphics[width=1.0\linewidth]{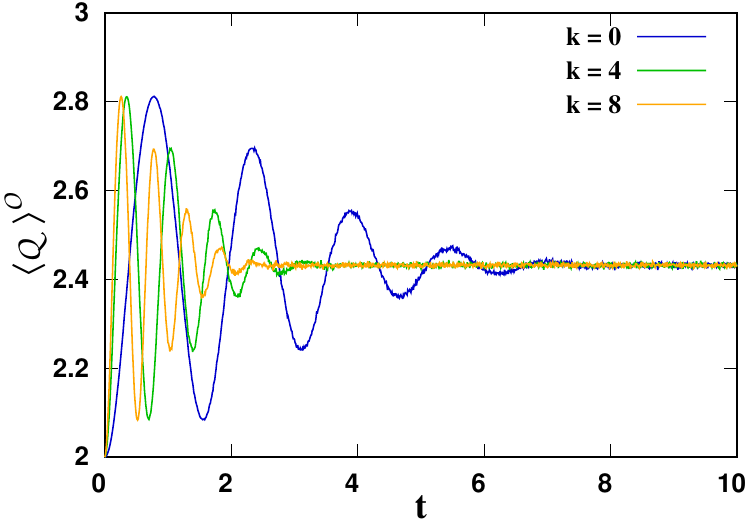}
\caption{(Color online.) The quenched averaged value of the maximal violation of local realism, $\langle \mathcal{Q} \rangle ^O$ (vertical axis) for different values of $k=0,4$ and $8$ against time, $t$ (horizontal axis), where $k$ denotes initial $k$th Fock state in the CV sector. Here, $\bar{\lambda} = 1.0, \sigma_{\lambda} = 0.1$. Saturation value of $\langle \mathcal{Q}^O \rangle_{\text{sat}}$ is $\approx 2.43$, and is independent of $k$. Both axes are dimensionless.}
\label{fig:n=2} 
\end{figure}
As an example, we choose three sets of system parameters, $k = 0, 4, 8$, where $\langle \lambda \rangle = 1.0$, and $\sigma_{\lambda} = 0.1$  and plot their corresponding  $\langle \mathcal{Q}\rangle^{O}$ for various times. In each of the cases, we observe the following: 

$(1)$ The quenched averaged maximal violation after some initial oscillations saturate. 
Irrespective of $k$ values, we, interestingly, find that the  quenched averaged  violation, $\langle \mathcal{Q}\rangle^{O}$, saturates to $ \approx 2.43$, see Fig. \ref{fig:n=2}. The fact that the measurement settings can be tuned based on the particular disorder realization $\lambda$ generates an atypical consequence. The saturated quenched averaged violation of Bell inequality  $\langle \mathcal{Q}\rangle^{O}_{\text{sat}} \approx 2.43$ remains invariant under different disorder strengths, $\sigma_\lambda$. 

$(2)$ Relative saturation times for different $k$ values have same pattern. In particular, we see that for higher $k$ values,  the saturation times, $t_{cr}$ are shorter than the one obtained with low $k$. This might be interpreted from the fact that the integrand in  Eq. \eqref{eq:QAk} oscillates with a higher frequency for large $k$ values compared to the one with low $k$ values. The integration  results in the cancellation of these fast oscillating terms, thereby leading to a quicker saturation for larger $k$ values.

$(3)$ In addition, our investigation also reveals that the saturation times can be shorten with the increase of $\bar{\lambda}$ and $\sigma_\lambda$.

\begin{figure}
\includegraphics[width=1.0\linewidth]{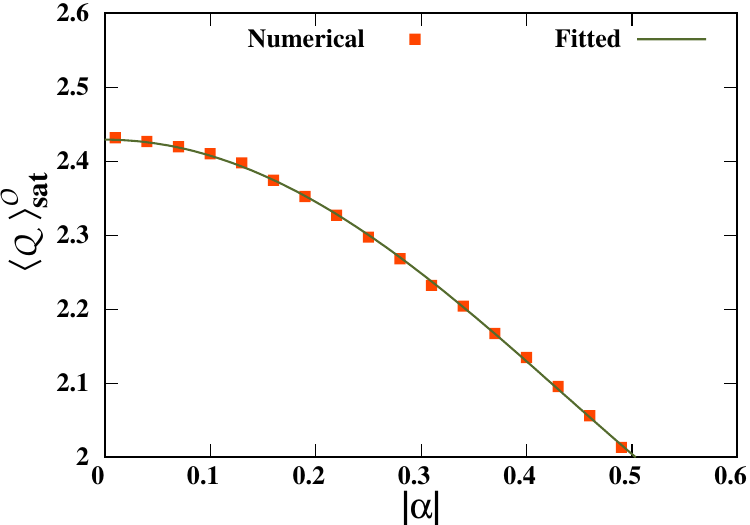}
\caption{(Color online.) The saturation value of the quenched averaged violation of Bell inequality, $\langle\mathcal{Q}\rangle^{O}_{\text{sat}}$ (ordinate)  with oracle measurement strategy against displacement of coherent state of the initial hybrid state,  $|\alpha|$ (abscissa). $\langle\mathcal{Q}\rangle^{O}_{\text{sat}}$ decreases with $|\alpha|$ till some critical value $|\alpha|_{cr}=0.51$ beyond which the violation vanishes. Both axes are dimensionless. }
\label{fig::oracle_saturation} 
\end{figure}

We carry out a similar analysis for  other choices of the initial state. An initial product state with the field state being in the coherent state and the atom being in the excited state, see Eq.[\ref{eq::coherent} ]. Note that all the other choices of initial states considered in this manuscript produce qualitatively similar results. 
 \begin{figure}
\includegraphics[width=1.0\linewidth]{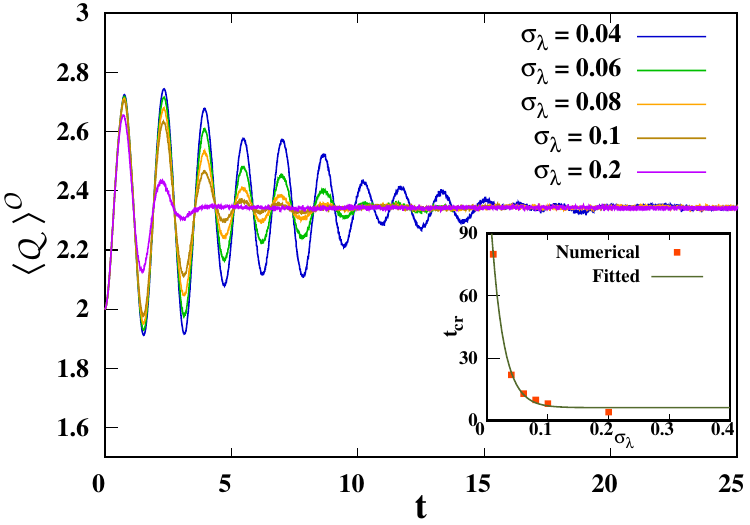}
 \caption{(Color online.) Dynamics of $\langle\mathcal{Q}\rangle^{O}$ (ordinate) vs $t$ (abscissa). The initial CV part of the hybrid system is in a coherent state with $|\alpha| = 0.2$ and mean interaction strength is $\bar{\lambda} = 1.0$. We take different values of standard deviation in $\lambda$ denoted by $\sigma_{\lambda} = 0.04,0.06,0.08,0.1$ and $0.2$.  (Inset) The critical time, $t_{cr}$ (vertical axis), i.e., the time when $\langle\mathcal{Q}\rangle^{O}$  reaches the  saturation value $\langle\mathcal{Q}\rangle^{O}_{\text{sat}}$ against  strength of the disorder, $\sigma_{\lambda}$ (horizontal axis). The number of realizations considered to perform quenched averaging is $7000$. All the axes are dimensionless.}
\label{fig:compare_disorder_channel} 
\end{figure}
Nevertheless, here also, the time dynamics of the quenched averaged maximal violation of Bell inequality observe a similar trend to when the initial state was taken to be in any of the Fock states -- initial oscillations followed by saturation, as depicted in Fig. \ref{fig:compare_disorder_channel} for various choice of disorder parameters. However, unlike in the previous case where the saturated value of the quenched averaged maximal violation is independent of $k$, here the saturated value $\langle\mathcal{Q}\rangle^{O}_{\text{sat}}$ decreases with the increase of $|\alpha|$, see Fig. \ref{fig::oracle_saturation}. In particular, we find a critical value of  $|\alpha|$=$|\alpha_{cr}| \approx 0.51$ such that $\langle\mathcal{Q}\rangle^{O}_{\text{sat}}=0$,for $|\alpha|>|\alpha_{cr}|$. By employing the $\chi$-square curve fitting, we realize that the functional form of $\langle\mathcal{Q}\rangle^{O}_{\text{sat}}$ with $|\alpha|$ to be $ a + b |\alpha|^2 + c|\alpha|^4$ with $a = 2.43, b = -2.18$ and $c = 1.91$ having maximum $1.92\%$  errors in parameters, as shown in Fig.~\ref{fig::oracle_saturation}.

We also investigate how the critical time $t_{cr}$ for $\langle\mathcal{Q}\rangle^{O}$ to saturate, varies with the disorder parameters and find it to be decreasing with increasing $\sigma_{\lambda}$ before eventually saturating at a constant value (see Fig.~\ref{fig:compare_disorder_channel}). Further by using the $\chi$-square curve fitting, the functional form of $t_{cr}$ with $\sigma_{\lambda}$ is found to be $ b + c \exp(-d(\sigma_{\lambda}- 0.01))$ with $b = 6.20, c = 73.40$ and $d = 49.30$ having maximum $10.0\%$  errors in parameters. Interestingly, like in the previous case where the initial CV state was taken to be in a Fock state, the saturated quenched averaged  violation of Bell inequality is independent of the disorder strength $\sigma_{\lambda}$.

 \subsection{Realistic measurement strategy} 
In the previous situation, we assumed that the experimentalist can adjust the measurement settings depending on the particular realization of the disorder.  
 In a more realistic scenario, the experimentalist has information only about  some global properties of the disorder, say a few moments or the range of the distribution. For example, if the disorder is Gaussian,  the measurement strategy is fixed only on the knowledge of the mean $\bar{\lambda}$ and the standard deviation $\sigma_{\lambda}$.  
One reasonable strategy that one can employ is to fix the measurement settings that are optimal for the mean of the disorder distribution $\bar{\lambda}$ for all cases. 
This choice is irrespective of the standard deviation $\sigma_\lambda$. In absence of realization-dependent tuning of measurement settings, this strategy seems to be the only rational one and, therefore, we consider this for our investigation.
 
We now compute the quenched averaged violation of local realism when the measurement settings are fixed to the direction that is optimal for the mean value $\bar{\lambda}$ of the disorder. For this, we first rewrite Eq. \eqref{eq:QAoracle} in a form to explicitly illustrate how the measurement settings enter the Bell expression as
\begin{equation}
     \mathcal{Q}^O(\lambda) =
      \sqrt{||\mathit{T}_{\lambda}.C_{\lambda}||^2 + ||\mathit{T}_{\lambda}.{C}'_{\lambda}||^2 },
      \label{eq:dissetting}
\end{equation}
where $C_{\lambda}$ and ${C}'_{\lambda}$ are two eigenvectors of $T_{\lambda}$ corresponding to the two largest eigenvalues in $\mathbb{R}_{3}$. The norm of a vector $X$ is given by $||X||= \sqrt{X^{\dagger}X}$. Physically, $C_{\lambda}$ and ${C}'_{\lambda}$ come from tuning the measurement direction optimally for a given realization $\lambda$, and hence denote the measurement settings. When the measurement direction is fixed to the one which is optimal for $\bar{\lambda}$, for a given realization $\lambda$, we get
\begin{eqnarray}
  && \mathcal{Q}^{\mathit{P}}(\lambda) = \sqrt{||\mathit{T}_{\lambda}.C_{\bar{\lambda}}||^2 + ||\mathit{T}_{\lambda}.{C}'_{\bar{\lambda}}||^2 } .
 \end{eqnarray}
Similar to the previous section, we take the quenched averaged value over different realizations is given by $\langle \mathcal{Q}\rangle^{P} = \int\mathcal{Q}^{P}(\lambda)P(\lambda)d\lambda.$  Note that we clearly have $\mathcal{Q}^{\mathit{P}}(\lambda) \leq \mathcal{Q}^{O}(\lambda)$ which directly translates to $\langle \mathcal{Q}\rangle^{P} \leq \langle \mathcal{Q}\rangle^{O}$. The equality holds when $P(\lambda) = \delta(\lambda-\bar{\lambda})$.
 
\begin{figure}
\includegraphics[width=1.0\linewidth]{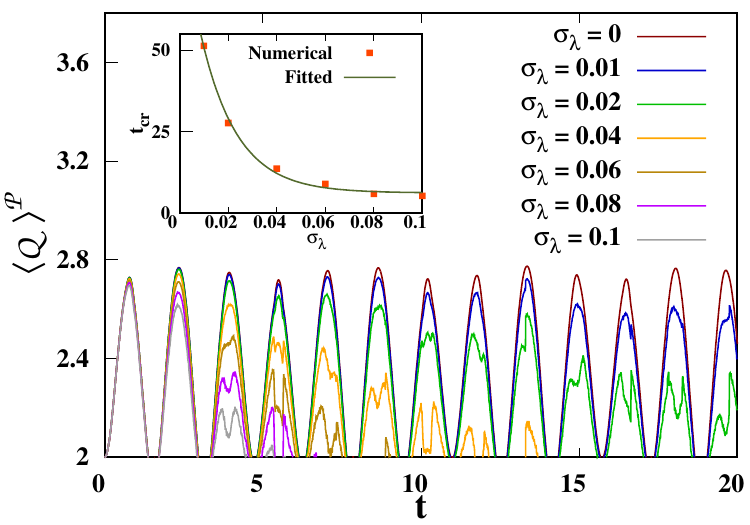}
\caption{(Color online.) Comparing the quenched averaged value of the violation of Bell inequalities $\langle\mathcal{Q}\rangle^{P}$ (ordinate) with time $t$ (abscissa) in realistic measurement settings  for $7000$ realizations. (Inset) The critical time $t_{cr}$ (ordinate) until which it is capable of detecting a violation of Bell inequality after quenched averaging vs \(\sigma_\lambda\) (abscissa).
The initial state of the mode is taken to be in a coherent state with $|\alpha|=0.2$. Further, by using the $\chi$-square curve fitting, we find the functional form of $t_{cr}$ with $\sigma_{\lambda}$ to be $ b + c \exp(-d(\sigma_{\lambda}- 0.01))$ ($b = 6.24, c = 44.65$ and $d = 66.79$ having maximum $10.0\%$  errors in parameters). All the axes are dimensionless.}
\label{fig:physical_disorder} 
\end{figure}

Realistic measurement scenario captures truly the essence of stochasticity in the system parameters and shows a drastically different dynamics with time. In the previous scenario, we found a saturation time of $\langle\mathcal{Q}\rangle^{O}$. In this realistic scenario, as the measurement settings is fixed to give maximal violation for $\bar{\lambda}$, for realizations that are more deviated from $\bar{\lambda}$ will contribute more non-optimally towards the quenched averaged value. For this reason, we are unable to find a situation where $\langle \mathcal{Q}\rangle^{P} > 2 $ through the whole time. We observe that the violation $\langle \mathcal{Q} \rangle ^P$ obtained for the disordered situation is enveloped by the violation of Bell inequality by the ordered system ($\sigma_{\lambda}=0$). 
Moreover, for any value of $\sigma_{\lambda}>0$, we find that $\langle \mathcal{Q}\rangle^{P}$  oscillates until a critical time $t_{cr}$ , such that for $t>t_{cr}$ , the value of  $\langle \mathcal{Q}\rangle^{P}$ goes below $2$. Hence, the realistic measurement settings with stochastic evolution parameter can only catch the  violation of local realism up to a certain time $t_{cr}$ for a given $\sigma_{\lambda}$ (as depicted in  Fig.~\ref{fig:physical_disorder}).


\section{Conclusion}
\label{sec:conclusion}
The detection of nonlocality and nonclassicalaity in continuous variable (CV) quantum systems has received a great deal of attention with the advancement of photonic quantum information technologies.
On the other hand, in discrete systems,  violations of local realism have been thoroughly studied over the years as a component of experimentally viable entanglement detection criteria and as a means of addressing the fundamental question of the completeness of quantum mechanics posed by Einstein-\textcolor{black}{Podolsky} and Rosen.
 Like violation of  Bell inequalities,  negative Wigner volume corresponding to quantum systems indicates correlations present in the quantum systems which cannot be classically simulated, although there is no strong connection between negative Wigner volume and violation of Bell inequalities. In this work, we exploited hybrid measurements made up of spin and pseudospin measurements to compute Bell correlations for detecting entanglement in hybrid systems composed of both discrete and CV systems. 

We investigated the maximal violation of Bell inequalities and  Wigner negativity in the dynamics of the hybrid system produced by the Jaynes-Cummings interaction between  finite- and the infinite-dimensional CV systems.  
 In particular, starting with the product, classically correlated, and entangled states, we investigated the pattern in the violation of Bell inequalities \ph{and entanglement} of the evolved state and discovered how the violation depends on the system characteristics, notably those involved in CV systems. \ph{Although for the CHSH inequality, typically nonlocality and entanglement have a one-to-one connection, our work does not exhibit so possibly due to the incapability of implementing the globally optimal measurement in the CV subsystem. However, the hybrid measurement proposed in our framework demonstrates a sufficient criterion for certifying the nonlocality of dynamical states of hybrid quantum systems. }

Deviating from the ideal scenario, we assumed that the interaction strength between the CV and the discrete systems has a fluctuation around a mean value. Following the oracle measurement strategy in which the measurement settings in the Bell expression can be tuned depending on the realization of a particular choice of a parameter, we demonstrated that the quenched averaged value of the maximal violation
of the Bell inequality saturates with time to a particular value,  independent of the fluctuation strength of the interaction parameter and with the moderate values of the displacement strength in the coherent state. However,  in the ordered case, it is always oscillating with time.  Therefore, the clear advantage of the disordered systems over the ordered case emerges in presence of moderate displacement value where the saturated value is finite for the entire time duration after the initial oscillation although  the evolved state does not show violation with the increase of the displacement parameter. Moreover, we presented a more realistic measurement strategy where the measurement direction is fixed to the direction of the maximal violation of Bell inequality for the mean value of interaction strength. In this case, the violation of Bell inequality in the dynamical state does not saturate to a particular value --  the violation decreases with time, which depends primarily on the standard deviation of the interaction strength, and it finally vanishes. \ph{The formalism presented in this work paves the way to demonstrate the nonlocality in hybrid quantum systems, with a possibility of near-term realization in experiments.}

\section*{acknowledgements}

We thank Ganesh for fruitful discussions. PH, RB, and ASD acknowledge the support from the Interdisciplinary Cyber Physical Systems (ICPS) program of the Department of Science and Technology (DST), India, Grant No.: DST/ICPS/QuST/Theme- 1/2019/23. This research was supported in part by the ‘INFOSYS scholarship for senior students’. We  acknowledge the use of \href{https://github.com/titaschanda/QIClib}{QIClib} -- a modern C++ library for general purpose quantum information processing and quantum computing (\url{https://titaschanda.github.io/QIClib}), and the cluster computing facility at the Harish-Chandra Research Institute. 
 This work has been partly supported by the Hong Kong Research Grant Council (RGC) through grant 17300918.

\appendix

\textcolor{black}{
\section{Proposal for the experimental realization of pseudospin measurements}
\label{app:proposal}}

\ph{The experimental realization of pseudospin measurements has been proposed \cite{Gardiner1997,Chen2002} by engineering Jaynes-Cummings interaction between the cavity mode with a sequence of $N$ two-level atoms. Specifically, the interaction is given by
\begin{eqnarray}
    H = g a^\dagger \sum_{i=1}^N \sigma^-_i
\end{eqnarray}
with $\sigma^-_i$ being the anihilation operator in the spin-$1/2$ system. During the interaction time, the pseudospin observable, $S_i^q$  on the cavity field fully develops into the corresponding observable of atoms, $M_S$ in the asymptotic limit, i.e.,
\begin{eqnarray}
    \lim_{N\to \infty}U_N^\dagger(t_I)S^q U_N(t_I) = \mathbb{I}_N\otimes M_S,
\end{eqnarray} 
where $U_N(t_I)=\exp(-iHt_I)$. Now the expectation value of pseudospin operator in the CV mode $\ket{\psi(0)}_\text{field}$ can be calculated as
\begin{widetext}
\begin{eqnarray}
_\text{field}\bra{\psi(0)}S^q\ket{\psi(0)}_\text{field} = \lim_{N\to\infty} {_\text{atom}}\bra{\psi(0)}^{\otimes N}_\text{field}\bra{\psi(0)}U_N(-t_I)(\mathbb{I}_N\otimes M_S)U_N^\dagger(-t_I)\ket{\psi(0)}_\text{field}\ket{\psi(0)}_\text{atom}^{\otimes N}.
\end{eqnarray}
\end{widetext}
Although $N\to\infty$ is impossible to achieve experimentally, high accuracy can be achieved with a finite number of atoms.}

\vspace{0.3cm}

\ph{\section{Correlation between Bell inequality violation and Wigner negativity volume}
\label{sec:app}

To analyze the correlation between $(|\mathcal{B}|_{\max}-2)$ and $\mathcal{V}_n$,  we plot 
\begin{figure}
\includegraphics[width=1.0\linewidth]{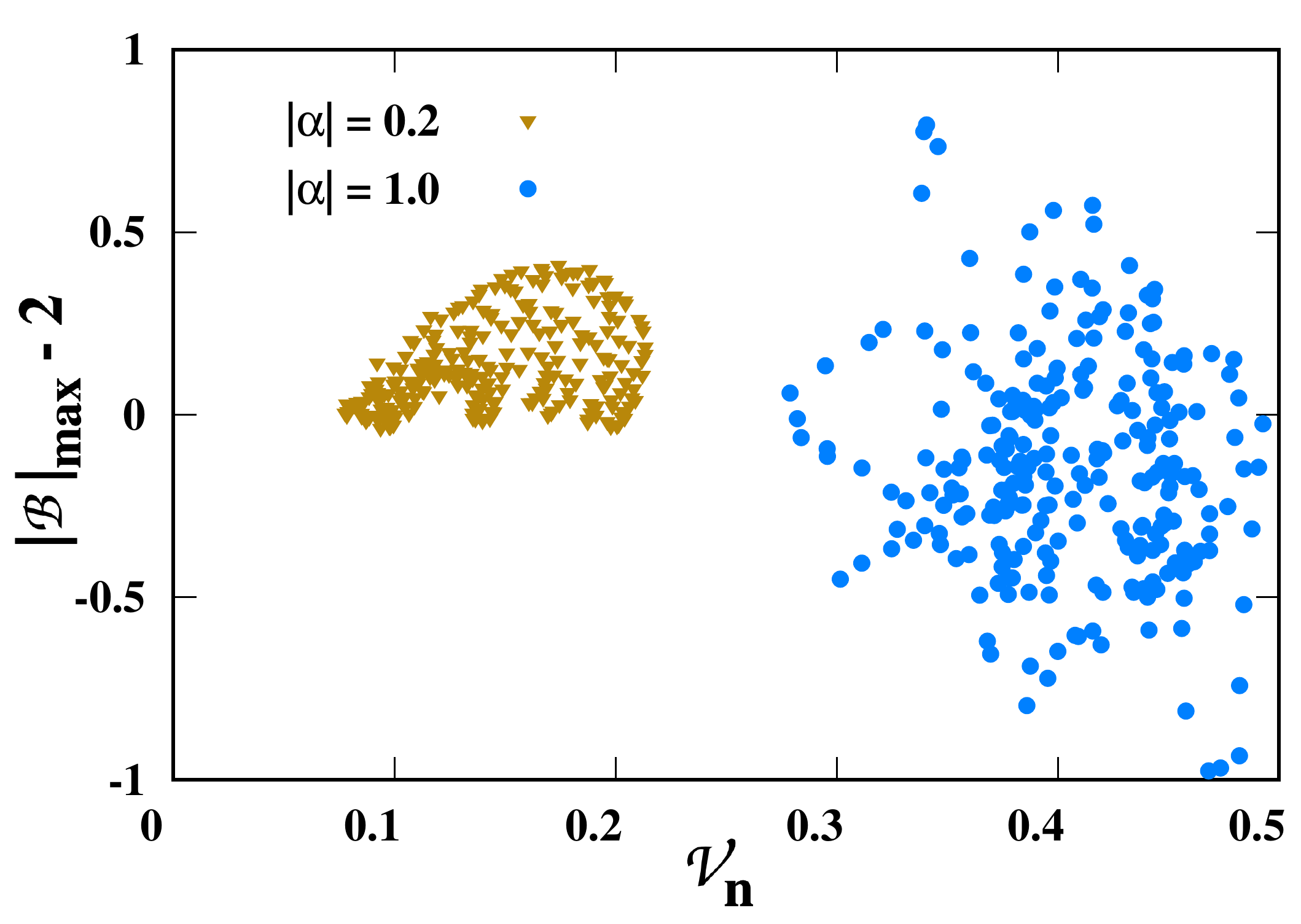}
\caption{(Color online.) Violation of Bell inequality, $(|\mathcal{B}|_{\max}-2)$ (vertical axis) is plotted against Wigner negative volume, $\mathcal{V}_n$ when the initial state is hybrid cat state for $|\alpha|=0.2$ (golden yellow triangles) and $|\alpha|=1.0$ (blue filled circles). All the axes are dimensionless.}
\label{fig:correl} 
\end{figure}
these two quantities in Fig.~\ref{fig:correl}. From the figure, we can see that as the value of $|\alpha|$ for the hybrid cat state (see Eq.~\eqref{eq:catgen}) increases, the region of nonviolation of Bell inequality increases. There exist regions where $|\mathcal{B}|_{\max} - 2\leq0$, i.e., there is no violation of Bell inequality, but we obtain nonzero $\mathcal{V}_n$. Moreover, the scattered nature of the plot depicts that there possibly does not exist any strong correlation between these two quantities.

To quantify the correlation between these two quantities, we have now calculated the Pearson correlation coefficient (PCC), $r_{xy}$ between two sets of data, denoted as $\{x_i=(|\mathcal{B}|_{\max}-2)_i, y_i=(\mathcal{V}_{n})_i\}_i$. Here, $r_{xy}$ is defined as
\begin{eqnarray}
    r_{xy}={\frac {\sum _{i}(x_{i}-{\bar {x}})(y_{i}-{\bar {y}})}{{\sqrt {\sum _{i}(x_{i}-{\bar {x}})^{2}}}{\sqrt {\sum _{i}(y_{i}-{\bar {y}})^{2}}}}},
\end{eqnarray}
where $\bar x$ and $\bar y$ denotes the mean value of the dataset $\{x_i\}_i$ and $\{y_i\}_i$ respectively. The value of PCC is found to be
\begin{eqnarray}
    r_{xy} = 
    \begin{cases}
        0.0016 & \text{for $|\alpha| = 0.2$},\\
        -0.0020 & \text{for $|\alpha| = 1.0$}.
    \end{cases}
\end{eqnarray}
Moreover, from the scattering nature of the plot, it is evident that the higher-order correlation coefficients must be vanishingly small. Therefore, the following observation immediately captures the poor correlation between Wigner negativity volume and nonlocality as expected.}

\bibliographystyle{apsrev4-1}
	\bibliography{bib.bib}

\end{document}